\documentclass[prb,aps,superscriptaddress,twocolumn,showpacs]{revtex4}
\usepackage{feynmp}
\usepackage{amsmath}
\usepackage{amssymb}
\usepackage{epsfig}
\usepackage{graphicx}
\usepackage{subfigure}
\usepackage{hyperref}
\usepackage{bm}
\usepackage{color}
\usepackage{longtable}
\usepackage{booktabs}
\usepackage{multirow}
\def\bea{\begin{eqnarray}}
\def\eea{\end{eqnarray}}
\def\l{\left}
\def\r{\right}

\def\nn{\nonumber}
\def\Eq#1{Eq.~(\ref{#1})}
\def\Eqs#1{Eqs.~(\ref{#1})}
\def\Fig#1{Fig.~\ref{#1}}
\def\abs#1{\left|#1\right|}
\def\xk#1{\left(#1\right)}
\def\zk#1{\left[#1\right]}
\def\dk#1{\left\{#1\right\}}

\def\Re{{\rm Re}}
\def\Im{{\rm Im}}
\def\sgn{{\rm sgn}}
\def\trace#1{{\rm Tr}\left[#1\right]}
\def\Det#1{{\rm Det}\left[#1\right]}

\def\pa{\partial}
\newcommand{\s}{{\sigma}}

\newcommand{\de}{\delta}
\newcommand{\De}{\Delta}
\newcommand{\ep}{\epsilon}
\newcommand{\ga}{\gamma}
\newcommand{\Ga}{\Gamma}

\newcommand{\La}{\Lambda}
\newcommand{\om}{\omega}

\renewcommand{\th}{\theta}

\renewcommand{\v}[1]{{\bf #1}}

\begin{document}

\title{Condition for the emergence of a bulk Fermi arc in disordered Dirac-fermion systems}

\author{Peng-Lu Zhao}
\altaffiliation{zpljdwlx@mail.ustc.edu.cn} \affiliation{Department
of Modern Physics, University of Science and Technology of China,
Hefei, Anhui 230026, P. R. China}
\author{An-Min Wang}
\affiliation{Department of Modern Physics, University of Science and
Technology of China, Hefei, Anhui 230026, P. R. China}
\author{Guo-Zhu Liu}
\affiliation{Department of Modern Physics, University of Science and
Technology of China, Hefei, Anhui 230026, P. R. China}

\begin{abstract}
We present a renormalization group analysis of the disorder effects
on the low-energy behaviors of two-dimensional tilted Dirac-fermion
systems, in which the fermions have two distinct orbitals unrelated
by any symmetry. Four types of disordered potential, two
interorbital and two intraorbital, are considered. If there is only
one type of interorbital disorder, the fermion-disorder scattering
induces logarithmic or power-law corrections to the fermion density
of states and specific heat. In contrast, the intraorbital disorder
can turn the system into a strongly disordered phase. In this
disordered phase, calculations based on self-consistent Born
approximation reveal that the Dirac point is destroyed and replaced
by a bulk Fermi arc. We also study the interplay of four types of
disorder, and find that the Dirac point can either remain intact or
give place to a Fermi arc. We obtain the condition for the emergence
of a Fermi arc in this case. Our results indicate that disorders can
result in rich low-energy properties of tilted Dirac fermions.
\end{abstract}


\maketitle

\section{Introduction \label{Sec:Into}}

Tilted Dirac/Weyl semimetal (SM), characterized by the tilting of
the conic spectrum of fermionic excitations \cite{Katayama06,
Kobayashi07, Goerbig08, Kobayashi08}, has attracted increasing
theoretical and experimental interest. For sufficiently large tilt,
the Fermi surface crossing the Dirac nodes becomes lines in two
dimensions \cite{Katayama06, Kobayashi07, Goerbig08, Kobayashi08}
and a surface in three dimensions \cite{Soluyanov2015}. Such a
system is usually called type-II Dirac/Weyl SM \cite{Soluyanov2015,
YSun2015, Koepernik2016, Autes2016}. The tilt-induced unusual Fermi
surface is found to produce a variety of novel phenomena, including
unconventional magnetic-optical response \cite{Proskurin15,
ZMYu2016,Tchoumakov2016,Udagawa2016}, magnetic breakdown
\cite{OBrien2016}, anomalous Hall effect \cite{Zyuzin16, Steiner17},
and anomalous Nernst and thermal Hall effects \cite{Ferreiros17,
Saha17}. Meanwhile, several scenarios have been proposed to realize
tilted Dirac/Weyl fermions in different regimes \cite{Katayama06,
Kobayashi07, Goerbig08, Kobayashi08, Soluyanov2015, Xu15, Xu17}.
Recent angle-resolved photoemission spectroscopy experiments
\cite{LHuang2016, CWang2016, MZYan17, Noh17, KNZhang17} have
reported evidence of their existence.

In previous works, the two degenerate states at Dirac point usually
refer to the spin components. In this case, the free tilted Dirac
fermions respect at least one of the time-reversal, spatial
inversion, and particle-hole symmetries, although the fundamental
Lorentz symmetry is always broken by the tilt. However, it is in
principle possible that the two degenerate states of fermions arise
from two distinct degrees of freedom that are not related by any
symmetry. For example, on the (001) surfaces of topological
crystalline insulators SnTe and Pb$_{1-x}$Sn$_x$Te, the two
components of Dirac fermions \cite{Zeljkovic2014} are made out of
the cation Sn/Pb orbital and the anion Te/Se orbital, respectively.
Similar features occur in some heavy fermion SMs due to the
hybridization of $f$- and $d$-bands \cite{ShenNonHbt, Dai}.

Disorder plays different roles in tilted Dirac/Weyl fermion systems
with and without symmetry constraints. If at least one of the
time-reversal, spatial inversion, and particle-hole symmetries is
respected, disorder can lead to several possible quantum phase
transitions. For instance, a compressible diffusive metal (CDM)
phase \cite{Fradkin1, Fradkin2, shindou-murakami, ominato-koshino,
goswami-chakravarty, herbut-disorder, brouwer, Syzranov15A, Roy14,
Pixley15, Syzranov15B}, in which the fermions acquire a finite
zero-energy disorder scatting rate $\ga_0$ and a finite zero-energy
density of states (DOS) $\rho\xk{0}$, is realized if some kind of
disorder is coupled to a single tilted Weyl cone \cite{Trescher,
Sikkenk}. Disorder may trigger a metal-insulator transition in
tilted Weyl fermion system \cite{XCXie2017}. Moreover, disorder is
predicted to drive a novel topological phase transition between
type-I and type-II SM states by reducing the topological mass
\cite{XCXie2017, Park2017}.

In case the Dirac/Weyl fermions have two components that are not
related by any symmetry, the disorder effect is still not well
studied. In Ref.~\cite{PapajNonH}, the authors considered one
special type of disorder that has never been studied before, and
showed that such disorder can destroy the Dirac point and replace it
with a bulk Fermi arc. Thus far, it remains unclear how other types
of disorder influence the low-energy behavior of the system.

In this paper, we study the disorder effects on 2D tilted Dirac
fermions that have two distinct orbitals. We only consider bilinear
fermion-disorder couplings based on the requirement that any single
type of disorder can exist alone without generating other types of
disorder under the renormalization group (RG) flow. In particular,
we will identify four types of disorder (see
Appendix~\ref{Sec:ApenRG} for a detailed analysis), two interorbital
and two intraorbital. After carrying out RG calculations, we find
that the RG flow of one type of interorbital disorder behaves as the
random gauge potential (RGP) widely studied in conventional 2D Dirac
SM \cite{Ludwig1994}, and that the other one resembles random mass
(RM) \cite{Ludwig1994}. RGP is unrenormalized owing to a local
time-independent gauge symmetry, and produces a stable non-Fermi
liquid (NFL) state in which the density of states and specific
heat receive power-law corrections. Different from RGP, RM is a
marginally irrelevant perturbation, leading to logarithmic
enhancement of DOS and specific heat. Neither RGP nor RM could
generate the bulk Fermi arc \cite{PapajNonH}. The RG behavior for
intraorbital disorder is analogous to the random scalar potential
(RSP) of conventional 2D Dirac SM \cite{Ludwig1994}. Both of the two
types of intraorbital disorder are marginally relevant and turn the
system into a strongly disordered phase \cite{PapajNonH}. We
calculate the fermion self-energy in such a disordered phase by
using the self-consistent Born approximation (SCBA), and obtain two
different scattering rates, corresponding to two different orbitals.
The difference in these two scattering rates gives rise to a bulk
Fermi arc, consistent with previous work \cite{PapajNonH}.

We also consider the case in which all the four types of disorder
coexist, and demonstrate that they all become marginally relevant at
low energies due to their interplay, which invariably drives the
system into a strongly disordered phase. We re-perform an SCBA
calculation, and find that the system might exhibit either stable
Dirac point or bulk Fermi arc, depending on the values of a number
of model parameters. We obtain the conditions for the emergence of
Fermi arc after computing the fermion self-energy functions. It is
interesting that two distinct orbitals may have exactly the same
disorder scattering rate, which prevents the emergence of Fermi arc.

The remainder of the paper is organized as follows. We present the
model Hamiltonian and derive the RG equations in
Sec.~\ref{Sec:Model}. The impact of each single type of disorder and
the interplay of all four types of disorder are analyzed in
Sec.~\ref{Sec:RGanalysis}. We analyze the conditions for Fermi arc
or Dirac point to exist in Sec.~\ref{Sec:Farc}. We summarize the
results and highlight possible future works in Sec.~\ref{Sec:Summ}.

\section{Model and flow equations \label{Sec:Model}}

As the starting point, we consider non-interacting tilted Dirac
fermions near one single Dirac cone described by the Hamiltonian
\cite{PapajNonH} \bea H_0(\v{p}) = \psi^\dagger(\v{p})
\xk{v_x p_x\s_z + v_y p_y\s_x+ w v_x
p_x\sigma_0}\psi(\v{p}),\label{EqfrHami} \eea where $\psi^{\dag} =
\xk{\psi_1^{\dag}, \psi_2^{\dag}}$ is a two-component fermion field,
$\sigma_0$ is the $2\times 2$ identity matrix, and $\sigma_i$
($i=x,y,z$) are the three Pauli matrices. We use $v_x$ and $v_y$ to
denote the fermion velocities along the $x$- and $y$-directions,
respectively. Without loss of generality, we choose $v_{x,y} > 0$. A
dimensionless tilting parameter $w$ is introduced along the $x$-axis.
For type-I Dirac fermions, the tilt is limited to the range of
$\abs{w}<1$, whereas $\abs{w} > 1$ for type-II. The point at which
$\abs{w}=1$ is called Lifshitz transition point, which separates
type-I from type-II Dirac fermions. The energy dispersion of
Hamiltonian \Eq{EqfrHami} is given by \bea {\cal E}_{\pm}\xk{\v p} =
wv_{x}p_{x}\pm\sqrt{v_{x}^{2}p_{x}^{2}+v_{y}^{2}p_{y}^{2}}.
\label{Eqdisp} \eea For type-II Dirac fermions, setting $E_{\pm}=0$
produces two Fermi lines, described by the relation: \bea
v_yp_{y}=\pm v_x p_x\sqrt{w^2-1}.\label{Eqferline} \eea Now the
Fermi surface is not point-like. To perform RG analysis, one can
either adapt the low-energy effective theory to model the fermionic
excitations near the Fermi surface \cite{SQShen}, or employ a
parameter regularization scheme \cite{YuLee} to ensure that the RG
transformations are scaled to the Fermi surface. For our purpose, we
restrict the tilt to the range of $\abs{w}<1$ to make the Fermi
surface closed. It is worth mentioning that, such type-I tilted
Dirac fermions can be realized in several materials, including the
(001) surface state of SnTe \cite{Ando, SodemannFu} and the organic
conductor $\alpha$-(BEDT-TTF)$_2$I$_3$ \cite{Katayama06,
Kobayashi07, Goerbig08, Kobayashi08, Bender84, IsobePRL}.

We now incorporate various disordered potentials into the
non-interacting Hamiltonian by adding the following fermion-disorder
coupling term \cite{Ludwig1994, Nersesyan95, Altland02,
Stauber2005PRB} \bea H_{\mathrm{dis}}(\v{x}) = \sum_{\ga}
A_{\ga}(\v{x})\psi^\dag(\v{x})\ga\psi(\v{x}), \eea where
$A_{\ga}(\mathbf{x})$ represents a given type of randomly
distributed potential and $\ga$ is a $2\times 2$ matrix. Here,
$A_{\ga}(\mathbf{x})$ is assumed to be a quenched, Gaussian white
noise potential, characterized by two features
\begin{eqnarray}
\langle A_{\ga}(\mathbf{x})\rangle = 0,\quad \langle
A_{\ga}(\mathbf{x})A_{\ga'}(\mathbf{x}')\rangle =\de_{\ga\ga'}
\Delta_{\ga} \delta^2(\mathbf{x}-\mathbf{x}').\label{Eq_def_dis}
\end{eqnarray}
Dimensionless variance $\Delta_{\ga}$ measures the strength of
random potential.

Disorders are classified by the definition of $\ga$. For a Dirac
fermion system that respects at least one of time-reversal, spatial
inversion, and particle-hole symmetries, $\ga$ can be the identity
matrix or any of the three Pauli matrices \cite{Ludwig1994}. In
particular, $\ga =\s_0$ corresponds to RSP, $\ga = \s_{z}$ to RM,
and $\ga =(\s_{x},\s_{y})$ to the two components of an RGP. However,
for the tilted Dirac fermions under consideration, $\ga$ should be
chosen in a different way. As demonstrated in Ref.~\cite{Sikkenk},
the disorder defined by $\ga=\s_0$ and the one by $\ga=\s_z$ are not
self-closed in the RG analysis, because their coupling to Dirac
fermions inevitably generates different types of disorder. In this
paper, we choose to define intraorbital disorder by the matrix $\ga
= A\s_0 + B\s_z$, where $A$ and $B$ are two constants. According to
Appendix~\ref{Sec:ApenRG}, the self-closeness of the RG analysis
allows us to choose $\xk{A,B}=\xk{1/2,\pm 1/2}$. We identify these
two types of disorder as RSPs. For interorbital disorders, the
matrix can be formally written as $\ga = A'\s_x + B'\s_y$. To obtain
self-closed RG equations, we are forced to define $\ga=\s_x$ and
$\ga=\s_y$, which are called RGP and RM, respectively.

We now see that, there are only four possible types of disorder in
the model considered in this paper. Any one of them can exist alone.
However, to make our analysis more generic, we assume that they
exist simultaneously in the system and then study their effects by
carrying out RG calculations. The influence of each single disorder
can be easily obtained by removing the rest three types of disorder.
We will show that the coexistence of different types of disorders
leads to intriguing new physics.

The random potential $V(\mathbf{x})$ needs to be averaged.
Generically, this can be accomplished by means of three approaches,
including the supersymmetry technique \cite{Efetov82, Efetovbook},
the Keldysh formalism \cite{Kamenevbook}, and the replica trick
\cite{Edwards75, Lee85, Altlandbook}. At the perturbation level,
they are equivalent \cite{Gruzberg97, Lerner02, Syzranov15B}. Here,
we employ the replica trick, which has been most widely used in the
literature. To average over $V(\mathbf{x})$, we assume that the
spatial distribution of $A_{\ga}\xk{\v x}$ is Gaussian, described by
$P\zk{A_{\ga}} =\exp\zk{-\int d^2 \v x A_{\ga}\xk{\v
x}/(2\De_{\ga})}$. After applying the replica trick, we express the
total effective action in the Euclidean space as follows \bea S &=&
\int d\tau d^2\v x \psi^\dagger_m \zk{\partial_\tau - i(\sigma_z + w
\sigma_0)v_x \partial_x -iv_y \sigma_x\partial_y} \psi_m \nn \\&&
-\sum_{i}\frac{\Delta_{i}}{2} \int d\tau d\tau' d^2\v x
\xk{\psi^\dagger_m\s_i\psi_m}_x
\xk{\psi^\dagger_n\s_i\psi_n}_{x'}\label{Eqaction}, \eea where
$i=x,y,\pm$ are used to indicate the disorder type and the
parameters $\Delta_i$ represent the corresponding disorder coupling
constants. Moreover, $m$ and $n$ are replica indices, which are
summed over from $1$ to $R$ automatically. At the last step of the
calculation, the replica limit $R\rightarrow 0$ should be taken.
Although the replica limit may give rise to unphysical results in
some non-perturbative studies \cite{Sherrington75, Hemmen79,
Verbaarschot85}, it is well-defined in the perturbation RG analysis
\cite{Altlandbook}.

\begin{figure}
\centering \subfigure{\includegraphics[width=3in]{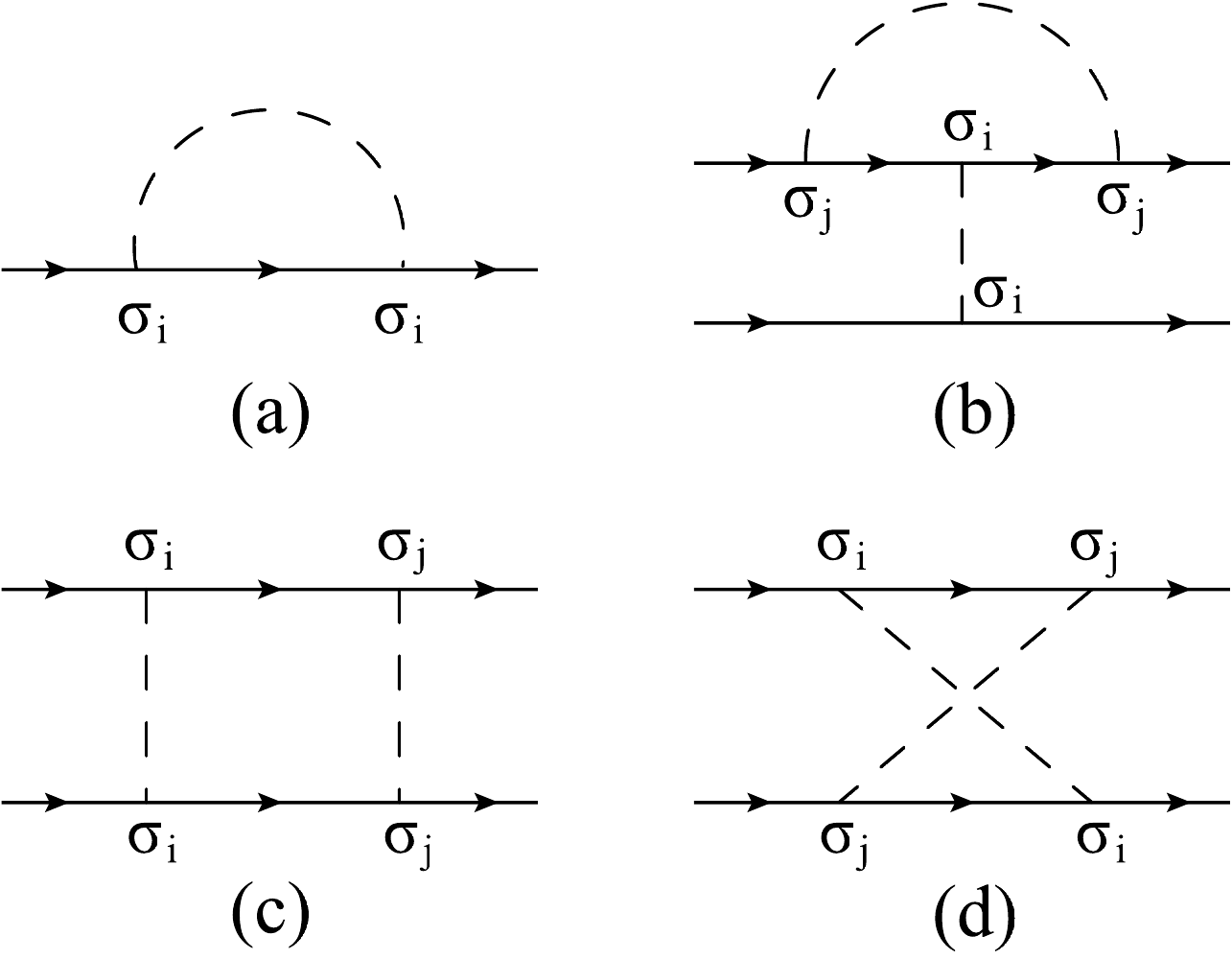}}
\caption{All the relevant one-loop Feynman diagrams. Solid line
represents free fermion propagator, and dashed line represents
disorder scattering.} \label{FigAll}
\end{figure}

We have completed the RG analysis of the action given by
\Eq{Eqaction}. Weak coupling expansion is adopted in the
perturbative calculation. All the relevant one-loop Feynman diagrams
are presented in \Fig{FigAll}. After integrating out the fast modes
defined within the momentum shell $e^{-\ell}\Lambda < |\mathbf{p}| <
\Lambda$, followed by RG transformations \cite{Shankar1994RMP}, we
obtain a number of coupled RG equations: \bea
\frac{dw}{d\ell}&=&\frac{\beta_-\xk{1+w}}{2}-\frac{\beta_+\xk{1-w}}{2}
-\xk{\beta_x+\beta_y}w \label{RGw}, \\
\frac{dv_x}{d\ell}&=&-\zk{\frac{\beta_++\beta_-}{2} +
\frac{\xk{\beta_x+\beta_y}\xk{1+w^2}}{1-w^2}}v_x, \label{RGvx} \\
\frac{dv_{y}}{d\ell} &=& -\frac{1}{2}\bigg[\frac{\xk{\beta_{x} +
\beta_{y}}}{1-w}+\frac{\beta_{+}}{1+w} +
\frac{\xk{\beta_{x}+\beta_{y}}}{1+w}\nn \\
&&+\frac{\beta_{-}}{1-w}\bigg]v_{y} \label{RGvy}, \\
\frac{d\beta_x}{d\ell}&=& \dfrac{\beta_y}{2}\zk{\beta_+
\xk{\frac{1-w}{1+w}}+\beta_-\xk{\frac{1+w}{1-w}}}, \label{RGbetax}
\\
\frac{d\beta_y}{d\ell} &=& -2\xk{\beta_y-\beta_x} \beta_y +
\dfrac{\beta_x}{2}\bigg[\beta_+\xk{\frac{1-w}{1+w}} \nn\\
&&+\beta_-\xk{\frac{1+w}{1-w}}\bigg]+\dfrac{\beta_+\beta_-}{2},
\label{RGbetay} \\
\frac{d\beta_{\pm}}{d\ell}&=&\frac{\mp w\beta _{\pm}^2}{1\pm w} +
\beta_{\pm}\bigg[\frac{\beta_{\mp}}{1\mp w}\mp \frac{2w
\left(\beta_x+\beta_y\right)}{1-w^2}\nn\\&&+\beta_x + 3\beta_y\bigg]
+ \frac{2\beta_x \beta_y \left(1\pm w\right)}{1\mp w},
\label{RGbetapm} \eea where \bea \beta_{i} \equiv
\frac{\Delta_{i}}{2\pi v_xv_y\sqrt{1-w^2}}\label{Eqeffcoup} \eea
represent the effective strength of fermion-disorder interaction.
The detailed RG calculational steps can be found in
Appendix~\ref{Sec:ApenRG}. In the next section, we analyze the
solutions to these RG equations and discuss the impact of various
random potentials on tilted Dirac fermions. We will first consider
each single type of disorder, and then their interplay.

\section{Disorder effects \label{Sec:RGanalysis}}

Our first aim in this section is to judge the relevance (or
irrelevance) of each type of disorder. For a marginal or an
irrelevant disorder, the perturbative RG is well under control in
the weak-coupling regime. The disorder effects can be examined by
directly computing the interaction corrections to some observable
quantities. For a relevant disorder, the perturbative RG method
breaks down as the disorder always flows to a strong coupling regime
at low energies. In this case, we will employ SCBA to calculate
disorder scattering rate and then to analyze the low-energy
properties of the strongly disordered phase.

\subsection{Random gauge potential \label{Sec:resultRGP}}

We first consider interorbital disorder scattering. When RGP exists
alone in the system, one can set $\beta_{y}=\beta_{+}=\beta_{-}=0$.
The RG equation of RGP parameter is simply \bea
\frac{d\beta_x}{d\ell}=0,\label{RGbetaxs} \eea which means that
$\beta_x$ is marginal at the one-loop level. Actually, this
conclusion is valid up to any order of loop expansion due to the
presence of a local gauge symmetry: \bea \psi_i \rightarrow \psi_i
e^{i\xi(y)}, \quad A(\v x)\rightarrow A(\v x)+ \pa_y \xi(y). \eea A
detailed proof of this un-renormalization is provided in
Appendix~\ref{Sec:ApenRGP}. In a (2+1)D untilted Dirac fermion
system, RGP is also marginal at any finite loop level, because the
time-independent gauge transformation guarantees that RGP is
un-renormalized \cite{Ludwig1994, Vafek08, Herbut08}. Such a
similarity provides further clue for the identification of the
disorder defined by $\ga=\s_x$ as RGP.

Since RGP is marginal, we will be able to obtain analytical
solutions for the fermion velocities and the tilt parameter. First,
we set \bea \beta_x\xk{E}=\beta_x\xk{\La},\label{EqSolux} \eea where
$\La$ is the UV cutoff that is determined by the bandwidth. $E = \La
e^{-\ell}$ is the low energy scale in which we are interested. The
constant $\beta_x\xk{\La}$ is the disorder strength defined at UV
cutoff $\La$. Hereafter, unless otherwise stated, the physical
quantities defined at $\La$ are taken as constants, but are regarded
as variables if defined at the varying energy $E$. Upon substituting
\Eq{EqSolux} into \Eq{RGw}, the tilt parameter has the form \bea
w\xk{E}=w\xk{\La}\xk{\frac{E}{\La}}^{\beta_x\xk{\La}},\label{EqSoluw}
\eea which vanishes as $E \rightarrow 0$. We then substitute
\Eqs{EqSolux} and (\ref{EqSoluw}) into \Eq{RGvx} and \Eq{RGvy}, and
obtain the following $E$-dependent velocities: \bea v_x\xk{E} &=&
\frac{v_x \xk{\La}\zk{1-w^2\xk{\La}} \xk{\La/E}^{\beta_x
\xk{\La}}}{\xk{\La/E}^{2\beta_x \xk{\La}}-w^2\xk{\La}},
\label{EqSoluvx} \\
v_y\xk{E} &=& \frac{v_y\xk{\La}\sqrt{1-w^2
\xk{\La}}}{\sqrt{\xk{\La/E}^{2\beta_x\xk{\La}}-w^2\xk{\La}}}
\label{EqSoluvy}. \eea In the low-energy region, we simplify these
expressions as power functions of momentum $k$, namely \bea
v_{x,y}\xk{E}|_{E \rightarrow 0} \sim E^{\beta_x\xk{\La}} \propto
k^{\eta_v}, \eea where $\eta_v = \beta_x\xk{\La}$. We can see that
$v_x$ and $v_y$ acquire the same finite anomalous dimension
$\eta_v$. A similar positive anomalous dimension has previously been
obtained in Ref.~\cite{WangLiu12} when studying the fermion velocity
renormalization of Dirac fermions in finite-density QED$_3$.
Moreover, this kind of fermion velocity renormalization is a special
property of several Dirac fermion systems, such as graphene
\cite{Vafek07, Son07, WangLiuNJP12, Sarma13, Kotov} and high-$T_{c}$
superconductors \cite{Kim97, Xu08, Huh08, Liu12, She15}. It may lead
to a number of unusual spectral and thermodynamic properties of
untilted Dirac fermions \cite{Vafek07, Son07, WangLiuNJP12, Kotov,
Kim97, Xu08, Huh08, Liu12, Sarma13, She15}. Here we show that this
phenomenon is also induced by RGP in the tilted Dirac fermion
system.

Next, we analyze the influence of marginal RGP on several important
quantities. The quasiparticle residue is defined as \bea Z_f =
\frac{1}{\abs{1-\frac{\pa}{\pa \om}
\Re{\Sigma}^{\rm{R}}\xk{\om}}_{\om\rightarrow 0}} \label{QPresdef},
\eea where $\Sigma^{\rm{R}}$ is the retarded fermion
self-energy. Since the electrons of $\s_z=\pm1$ orbitals are not
connected by any symmetry, their one-loop self-energy corrections
might be different, as can be seen from \Eq{Eqseenone} in
Appendix~\ref{Sec:ApenRG}. Accordingly, the residue could take
different values for the two orbitals. It is also possible to
compute the residue within the RG framework. Making use of the RG
solutions, we express the residue for $\s_z = \pm1$ orbitals in the
following form: \bea \frac{dZ_f^{\pm}}{d\ell}&=&-\frac{\beta_{x}
}{1\pm w} Z_f^{\pm}. \eea By substituting \Eqs{EqSolux} and
(\ref{EqSoluw}) into this equation, we find that: \bea
Z_f^{\pm}\xk{E}=\frac{1\pm w\xk{\La}}{\xk{\La/E}^{\beta_x\xk{\La}}
\pm w\xk{\La}}. \eea At low energies, the residue exhibits the same
energy dependence, namely \bea Z_f^{\pm}\xk{\om} \sim
\om^{\beta_x\xk{\La}}. \eea Here, we have simply replaced $E$ with
$\om$. Combining this result with \Eq{QPresdef}, we obtain \bea
\Re{\Sigma_{\rm{ii}}^{\rm R}}\xk{\omega} \sim
\om^{1-\beta_x\xk{\La}},\quad\xk{\rm i=1,2}. \eea Using the
Kramers-Kronig relation, we obtain the imaginary part of retarded
fermion self-energy: \bea \Im{\Sigma_{\rm{ii}}^{\rm R}}\xk{\omega}
\sim \om^{1-\beta_x\xk{\La}}. \eea Because $\beta_x\xk{\La} > 0$,
both of the two different orbitals display NFL-like low-energy
behaviors. Moreover, the difference in the residues $Z_f^{+}$ and
$Z_f^{-}$ vanishes at low energies.

According to the results of Appendix~\ref{Sec:Apenob}, the DOS and
specific heat of clean tilted Dirac fermions are given by \bea
\rho_0(E) &=& \frac{E}{\pi v_x v_y \left(1-w^2\right)^{3/2}},
\label{EqDOSclt}\\ C_{v}(T) &=& \frac{18\zeta(3)T^2}{\pi v_x v_y
\xk{1-w^2}^{3/2}}, \label{Eqshcl} \eea where $\zeta (x)$ is the
Riemann zeta function. Finite tilt tends to enhance DOS and specific
heat. As $\abs{w} \rightarrow 1$, the DOS formally diverges,
indicating the instability of the system. The disorder effects on
DOS and specific heat are embodied in the quantum corrections to the
fermion velocities and the tilt parameter. According to
\Eqs{RGw}-(\ref{RGvy}), although the velocity renormalization is not
dependent on the disorder type, the tilt parameter renormalization
is type sensitive. As a result, different types of disorder result
in different behaviors of DOS and specific heat.

When the RGP-induced corrections are taken into account, the fermion
DOS becomes \bea \rho\xk{E} &=&
\rho\xk{\La}\xk{\frac{E}{\La}}^{1-2\beta_x\xk{\La}} \sim
E^{1-2\beta_x\xk{\La}}. \label{EqDOSRGP} \eea Similarly, the
specific heat is altered by RGP to take the form \bea C_v\xk{T}&=&
C_v\xk{T_{\La}}\xk{\frac{T}{T_{\La}}}^{2-2\beta_x\xk{\La}} \sim
T^{2-2\beta_x\xk{\La}}, \label{EqshRGP} \eea where $T_{\La}$ is
certain fixed high temperature. Thus we see that RGP gives rise to
power-law enhancement of DOS and specific heat, characterized by two
tilt-independent exponents. This stems from the fact that RGP
reduces the tilt down to zero, which implies that the tilt becomes
irrelevant at low energies. In addition, the power-law enhancement
of DOS and specific heat can be interpreted as the emergence of
NFL-like behavior. The RGP-induced breakdown of FL theory
\cite{Pixley16A, Pixley16C, Roy16C, Fu17} has been studied in
several SM materials, including 2D Dirac SM \cite{Ludwig1994,
Altland02, Nersesyan95, Ostrovsky06, Foster12, Foster14, JingWang17}
and multi-Weyl SM \cite{JRWmultiW}. Analogous unusual behavior also
appears in Dirac SM with long-range correlated RM
\cite{Fedorenko12}.

\subsection{Random mass \label{Sec:resultRM}}

\begin{figure*}[htbp]
\centering
\subfigure{\includegraphics[width=3.2in]{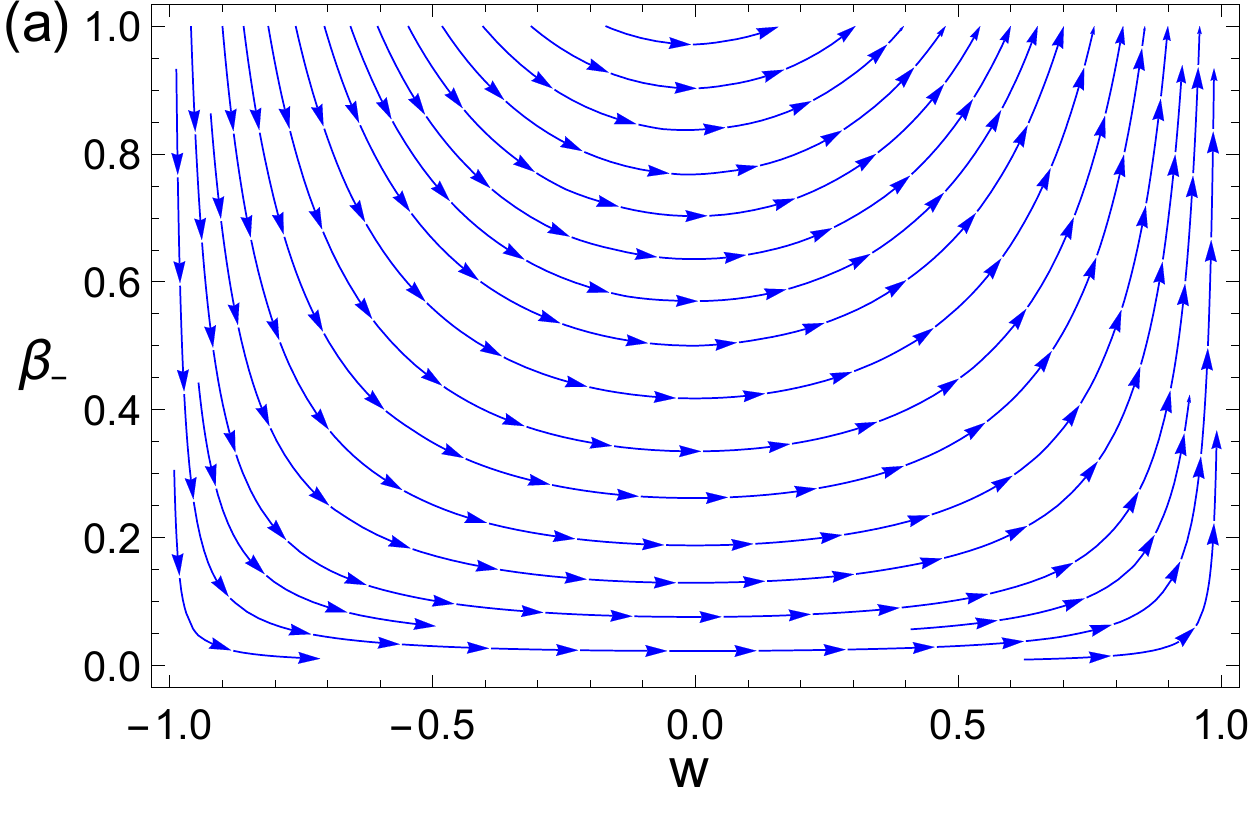}\label{FigFlpmm}}\hspace{0.4cm}
\subfigure{\includegraphics[width=3.2in]{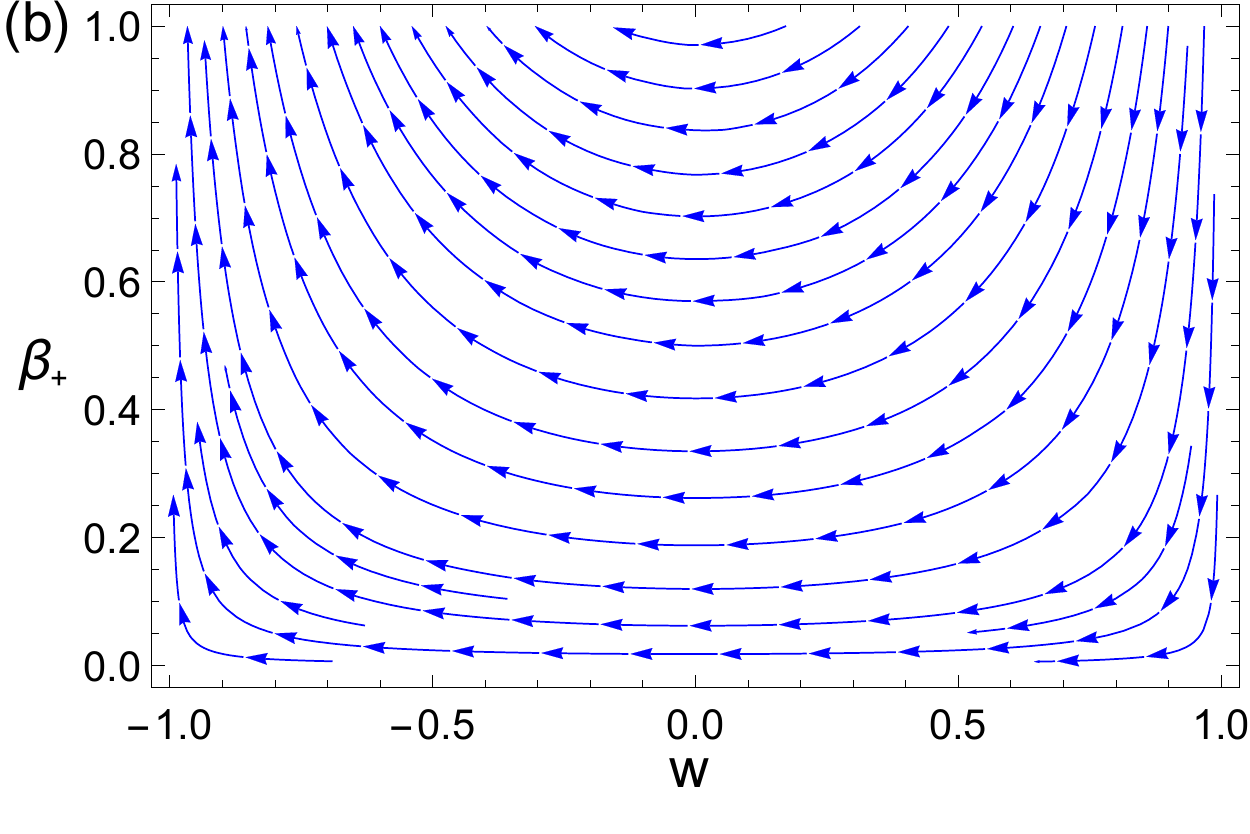}\label{FigFlpmp}}\vspace{-0.4cm}
\caption{RG flow diagram in (a) $w$-$\beta_-$ plane and (b)
$w$-$\beta_+$ plane.} \label{FigFlpm}
\end{figure*}

We now consider RM. According to \Eq{RGbetay}, when RM exists alone
its flow equation is given by \bea \frac{d \beta_y}{d\ell} =
-2\beta_y^2. \eea This equation has the following solution \bea
\beta_y\xk{E} = \frac{\beta_y\xk{\La}}{1+2\beta_y
\xk{\La}\ln\xk{\La/E}},\label{EqSoluy} \eea which approaches to zero
in the low-energy limit. Therefore, RM is a marginally irrelevant
perturbation. Repeating the RG steps performed in
Sec.~\ref{Sec:resultRGP}, we find the following solutions for the
tilt parameter and the fermion velocities: \bea w\xk{E} &=&
\frac{w\xk{\La}}{\sqrt{1 + 2\beta_y
\xk{\La} \ln\xk{\La/E}}},\label{EqSoluwRM} \\
v_x \xk{E} &=& \frac{\sqrt{1+2\beta_y\xk{\La}\ln\xk{\La/E}}}{1-w^2
\xk{\La}+2\beta_y\xk{\La}\ln\xk{\La/E}}\nn
\\ &&\times v_x\xk{\La}\zk{1-w^2\xk{\La}}, \label{EqSoluvxRM} \\
v_y\xk{E}&=&\frac{v_y\xk{\La}\sqrt{1-w^2\xk{\La}}} {\sqrt{1-w^2
\xk{\La}+2\beta_y\xk{\La}\ln\xk{\La/E}}}.\label{EqSoluvyRM} \eea All
of these three quantities go to zero logarithmically as $E
\rightarrow 0$. Interestingly, no anomalous dimension is generated.
It turns out that RM leads to weaker corrections to the properties
of tilted Dirac fermions than RGP. To confirm this, we now calculate
the quasiparticle residue. Based on \Eqs{QPresdef}, (\ref{EqSoluy}),
and (\ref{EqSoluwRM}), we obtain \bea Z_f^{\pm}\xk{\om}=\frac{1\pm
w\xk{\La}}{\sqrt{1+2\beta_y\xk{\La}\ln\xk{\La/\om}}\pm w\xk{\La}}.
\eea By replacing $E$ with $\om$, the energy dependence of
$Z_f^{\pm}$ is roughly given by \bea Z_f^{\pm}\xk{\om}\big|_{\om
\rightarrow 0} \propto \zk{\ln\xk{\frac{\La}{\om} }}^{-1/2} \eea as
$\om \rightarrow 0$. The real part of retarded fermion self-energy
is \bea \Re{\Sigma_{\rm{ii}}^{\rm R}}\xk{\omega} \sim \om
\zk{\ln\xk{\frac{\La}{\om}}}^{1/2}, \eea and the imaginary part is
\bea \Im{\Sigma_{\rm{ii}}^{\rm R}}\xk{\omega} \sim \om
\zk{\ln\xk{\frac{\La}{\om}}}^{-1/2}. \eea Thus RM also leads to
violation of FL theory, in a way similar to marginal Fermi liquid
(MFL) \cite{GonzalezMFFL, Varma02}. Again, the difference in the
residues of different orbitals becomes irrelevant at low energies.

After incorporating the corrections due to RM, the DOS and specific
heat become \bea \rho\xk{\om} &=& \rho\xk{\La}\xk{\frac{\om}{\La}}
\zk{1+2\beta_y\xk{\La}\ln\xk{\frac{\La}{\om}}} \nn\\ &\sim& \om \ln
\om, \nn\\
C_v\xk{T} &=& C_v\xk{T_{\La}} \xk{\frac{T}{T_{\La}}}^2
\zk{1+2\beta_y\xk{\La}\ln\xk{\frac{T_{\La}}{T}}} \nn\\ &\sim& T^2\ln
T. \eea The calculational details are shown in
Appendix~\ref{Sec:Apenob}. We conclude that the marginally
irrelevant RM only causes logarithmic enhancement of DOS and
specific heat. In analog to RGP, RM also suppress the tilt down to
zero at low energies, thus the tilt does not play an important role.
In the RG scheme, the influence of RM seems to be quite weak due to
its irrelevance. However, its effect might not be limited to such
logarithmic corrections. Rare region effects are believed to play an
important role \cite{NandkishoreRag, PixleyPRX,
Pixley2017,Gurarie2017, Holder2017} in the case of perturbatively
irrelevant disorder \cite{Fradkin2, Syzranov18}. Previous studies
suggest that when these effects are considered, the Dirac SM phase
could exist only in the ultra-clean limit \cite{NandkishoreRag}.
Moreover, due to rare region effects, the quantum critical region
between the SM and DM phases may become a sharp crossover, with the
putative critical point entirely avoided \cite{Pixley2017,
Gurarie2017, Holder2017}. However, rare region effects are
non-perturbative and thus cannot be studied by means of the perturbative
RG approach. We leave the rare region effects of marginally
irrelevant RM to future research.

\subsection{Random scalar potential \label{Sec:resultRSP}}

The results of the previous two subsections indicate that, although
RGP and RM strongly modify the low-energy properties of tilted Dirac
fermions, the system remains stable. In this subsection, we will
show that the role played by intraorbital disorder is entirely
different from interorbital disorder.

Suppose that only one electron orbital, either $\s_z=1$ or
$\s_z=-1$, is subjected to RSP. The RG equations of $w$ and
$\beta_+$ (or $\beta_-$) simplify to \bea \frac{d
\beta_{\pm}}{d\ell}&=&\frac{\mp w\beta _{\pm}^2 }{1\pm w},
\label{RGpm} \nn\\ \frac{dw}{d\ell}&=&\mp \frac{\beta _{\pm}\xk{1\mp
w}}{2}. \label{RGwpm} \eea We plot the RG flow diagrams for $\xk{w,
\beta_-}$ and $\xk{w, \beta_+}$ separately in \Fig{FigFlpmm} and
\Fig{FigFlpmp}. From these two diagrams, we see that disorder in the
$\s_z=-1$ orbital drives the tilt to $1$, whereas disorder in the
$\s_z=+1$ orbital drives the tilt to $-1$. As a result, the DOS of
disordered orbital becomes larger, according to \Eq{EqDOSclt}. The
enhanced DOS in turn increases the disorder strength. At ultra low
energies, the disorder strength formally flows to infinity. When
$w\rightarrow \pm 1$, the divergence of DOS and disorder strength
indicates that the system is no longer stable, but enters into a
strongly disordered phase. The perturbative RG approach cannot be
used to analyze the properties of this disordered phase. To obtain
the disorder scattering rate generated in the disordered phase, we
will employ the SCBA method to self-consistently calculate the
fermion self-energy. The case of $w\rightarrow 1$ has already been
analyzed by Papaj \emph{et al.} \cite{PapajNonH}. Here, we focus on
the case of $\s_z=+1$ orbital with $w\rightarrow -1$.

\begin{figure*}[htbp]
\centering
\subfigure{\includegraphics[width=3.05in]{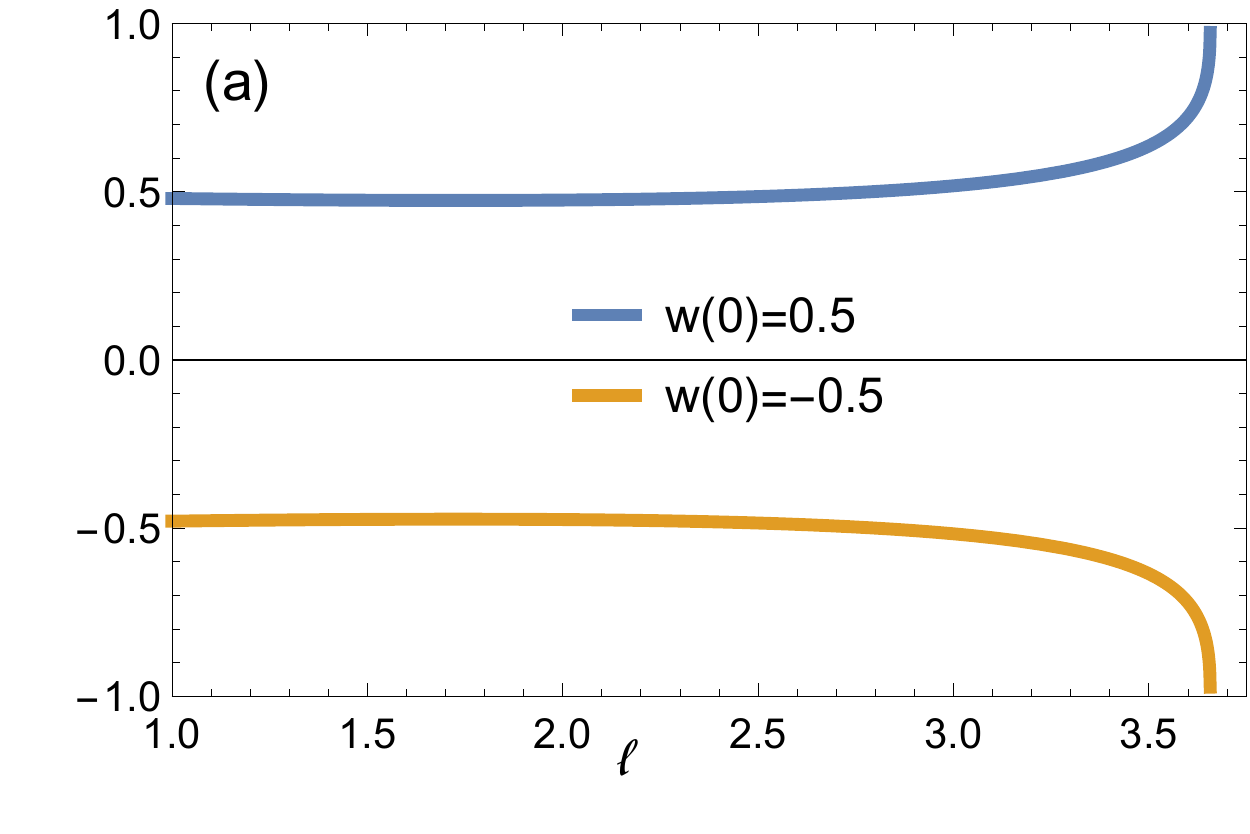}\label{FigCow}}\hspace{0.4cm}
\subfigure{\includegraphics[width=3in]{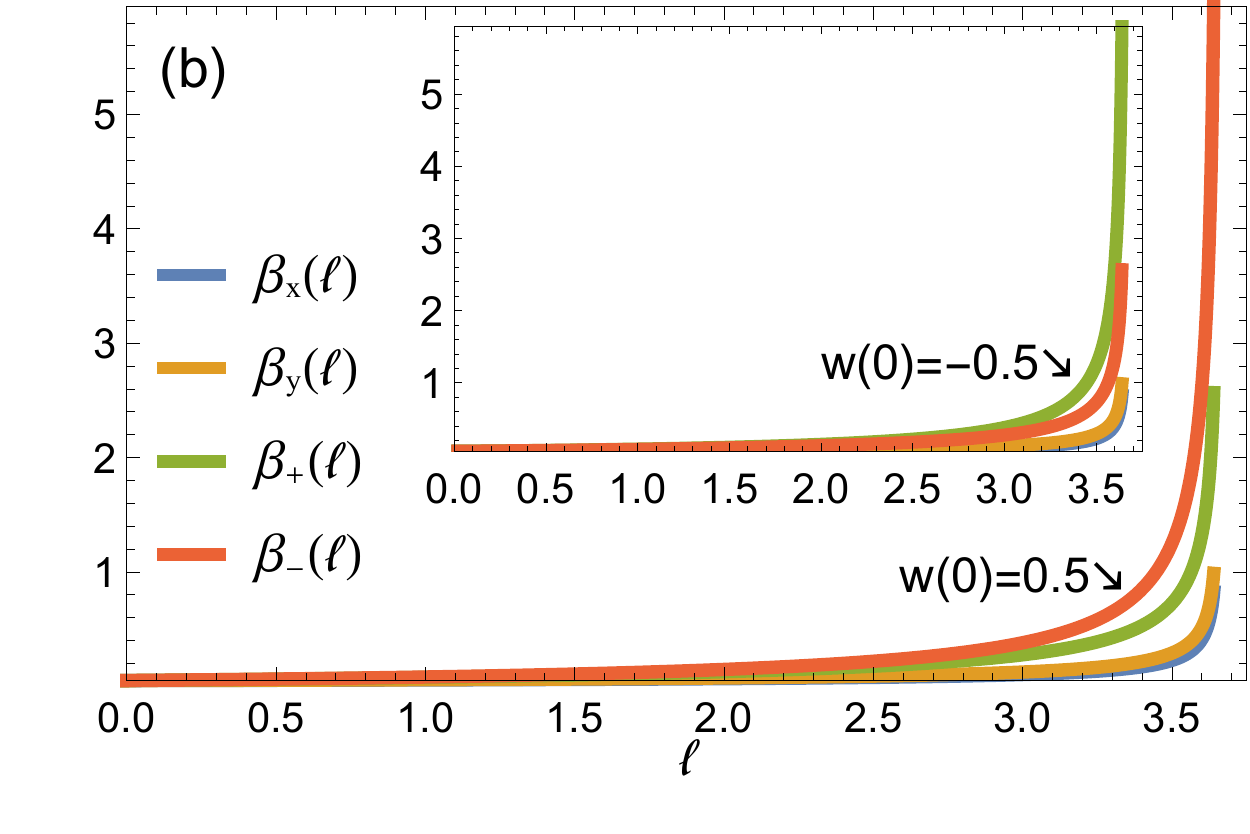}\label{FigCod}}\vspace{-0.4cm}
\caption{(a) Flow of $w\xk{\ell}$ at different initial values. (b)
Flowing behaviors of four disorder parameters obtained by choosing
two different values of $w\xk{0}$ in (a). Here, $\ell=\ln
\xk{\La/E}$ serves as the running scale. For (a), the initial value
of $\ell$ is chosen to visualize the jump in $w$. For (b), we take
$\beta_x(0) = \beta_y(0) = \beta_+(0) = \beta_-(0) = 0.05$.}
\label{FigCo}
\end{figure*}

Within SCBA scheme, the self-consistent equation for the fermion
self-energy in the $\s_z=+1$ orbital takes the following form \bea
\Sigma(\ep) &=& \Delta_+ \int''\frac{d^2k}{(2\pi)^2} \s_+
\frac{1}{\ep-H_0\xk{\v k}-\Sigma(\ep) } \s_+,\label{EqSCBAp} \eea
where the definition of $\int''$ is given by \Eq{Eqexplain}, in
Appendix~\ref{Sec:ApenSCBA}. As explained in
Appendix~\ref{Sec:ApenSCBA},
$\Sigma\xk{\ep}=\Sigma_{11}\xk{\ep}\s_+$ holds, and \Eq{EqSCBAp}
generates a self-coupled equation for $\Sigma_{11}\xk{\ep}$ of the
form: \bea \Sigma_{11}(\ep)&=&\frac{2\ep}{1-w}\mp
2i\La\sqrt{\frac{1+w}{1-w}} \nn\\&& \times
\exp\dk{\frac{-2\Sigma_{11}(\ep)\xk{1+w}}{\beta_+\zk{
\xk{1-w}\Sigma_{11}(\ep)-2\ep}}}, \label{EqCoSCBA11+p} \eea where
the upper and lower signs represent retarded and advanced
self-energy functions, respectively. The solution for
$\Sigma_{11}\xk{0}$ leads to the following constant \bea \Ga_1 =
2\La\sqrt{\frac{1+w}{1-w}} \exp\xk{-\frac{2}{\beta_+}
\frac{1+w}{1-w}}, \eea which defines a low energy scale. As the
energy decreases down to $\Ga_1$, the tilt approaches to $-1$, which
means the one-loop RG becomes invalid. For energies well beyond
$\Ga_1$, the self-energy can be calculated analytically. At
sufficiently high energies of $\abs{\ep}\gg \Ga_1$,
\Eq{EqCoSCBA11+p} can be solved by using the iterative method in
powers of $\beta_+$. At low energies $\abs{\ep}\ll \Ga_1$, the
solution can be obtained by taking a series expansion in powers of
$\ep$. Based on the calculations presented in
Appendix~\ref{Sec:ApenSCBA}, the self-energy is given by \bea
\Sigma_{11}\label{Eqsigma11s}\xk{\ep}= \left\{
\begin{array}{ll}
-\dfrac{4(1+w)\ep}{(1-w)^2\beta_+} \mp i\Gamma_1 &
\hbox{$\xk{|\ep|\ll\Gamma_1}$,} \\*[0.3cm]
\dfrac{-2\beta_+}{1+w}\zk{\pm i\pi\abs{\ep}/2 +\ep
\r.\\*[0.3cm]\times\l.\ln\xk{\La\sqrt{1-w^2}/\ep}} &
\hbox{$\xk{|\ep|\gg\Gamma_1}$.}
  \end{array}
\right. \eea This self-energy tells us that the electrons in the
$\s_z=+1$ orbital acquire a finite scattering rate at $\om=0$, i.e.,
\bea \ga_0=\abs{\rm{Im}\Sigma_{11}^R\xk{0}}=\Ga_1. \eea A fermion
system with a nonzero $\ga_0$ is often identified as a CDM
\cite{Fradkin1, Fradkin2, shindou-murakami, ominato-koshino,
goswami-chakravarty, herbut-disorder, brouwer, Syzranov15A, Roy14,
Pixley15, Syzranov15B}. However, this identification makes sense
only when the electrons of two orbitals have the same $\ga_0$. The
situation is different in our case, because the electrons in the
$\s_z=+1$ orbital have a nonzero $\ga_0=\Ga_1$ but those in the
$\s_z=-1$ orbital have $\ga_0 = 0$. As pointed out in
\cite{ShenNonH, KoziiNonH, PapajNonH}, the appearance of two
different scattering rates causes the original free Hamiltonian to
become non-Hermitian. As a result, the Dirac point disappears, and a
bulk Fermi arc emerges \cite{PapajNonH}.

\subsection{Coexistence of all four types of disorder\label{Sec:resultCo}}

In the last three subsections, each type of disorder is supposed to
exist individually in the system. But it happens in many realistic
materials that more than one types of disorder coexist. The
interplay of different types of disorder could give rise to much
richer low-energy behaviors \cite{Foster12, JingWang17}. According
to \Eq{RGbetapm}, the coexistence of an RGP and an RM generates an
RSP flow even if the system originally does not contain RSP. A more
complete analysis reveals that, the coexistence of any two types of
disorder invariably generates the rest ones. As a result, the system
are eventually driven to contain all the four types of disorder. To
analyze the properties of this situation, we need to analyze the
full set of RG equations presented in \Eqs{RGw}~-~(\ref{RGbetapm}).
Because the velocities are always reduced down to zero in the
low-energy limit, which can be inferred from \Eqs{RGvx} and
(\ref{RGvy}), we will pay special attention to the RG flows of $w$
and four disorder parameters.

By numerically solving the coupled RG equations, we obtain the
running behavior of $w$ and show the result in \Fig{FigCow}. The
$\ell$-dependence is sensitively determined by the sign of the
initial value of $w$. For a positive $w(0)$, $w\xk{\ell}$ approaches
to unity at a constant energy scale of $\ell_c=\ln\xk{\La/E_c}$.
However, $w \xk{\ell}$ flows to $-1$ at the same energy scale if
$w(0)$ is negative. The $\ell$-dependence of disorder parameters
obtained by starting from positive and negative $w(0)$ is plotted in
\Fig{FigCod}. The main figure presents the results for a positive
$w(0)$, and the inset for negative $w(0)$. We observe that, the
flowing behaviors of $\beta_x$ and $\beta_y$ are not affected by the
sign reversal of $w(0)$, whereas the flowing behaviors of $\beta_+$
and $\beta_-$ are interchanged. As shown in \Fig{FigCod}, $\beta_+ >
\beta_-$ when $w\rightarrow 1$ at a fixed scale, but $\beta_+ <
\beta_-$ when $w\rightarrow -1$. This is different from the case in
which only one single type of disorder exists. From the analysis of
Sec.~\ref{Sec:resultRSP}, it appears that a larger $\beta_-$ is
favored if $w\rightarrow 1$, and $w\rightarrow 1$ leads to a larger
$\beta_+$. The flip in this behavior is caused by the interplay
between different types of disorder.

Regardless of the subtle difference caused by different initial
values of $w$, we find that all the disorder parameters become
divergent at the same constant energy if $\abs w \rightarrow 1$.
Once again, the system is no longer stable and driven into a
strongly disordered phase. The new feature is that now four types of
disorder are present simultaneously. To gain further insight into
the strongly disordered phase, we will again make a SCBA analysis.
Now, the self-energy should be decomposed in the form
$$\Sigma\xk{\ep} = \Sigma_{11}\xk{\ep}\s_++\Sigma_{22}
\xk{\ep}\s_-.$$ Repeating the same calculational steps, we find the
following relation between $\Sigma_{11}\xk{\ep}$ and
$\Sigma_{22}\xk{\ep}$: \bea
\xk{\!\frac{\beta_x\!+\!\beta_y}{1\!+\!w}\!+\!
\frac{\beta_-}{1\!-\!w}\!}\!\Sigma_{11}(\ep) =
\xk{\!\frac{\beta_x\!+\!\beta_y}{1\!-\!w}\!+\!
\frac{\beta_+}{1\!+\!w}\!}\!\Sigma_{22}(\ep)\label{EqrelaS}. \nn
\\ \eea This is another important result caused by the interplay of
different disorders. This condition will be utilized to determine
under what circumstance a bulk Fermi arc is realized.

\begin{figure*}[htbp]
\centering
\subfigure{\includegraphics[width=3.0in]{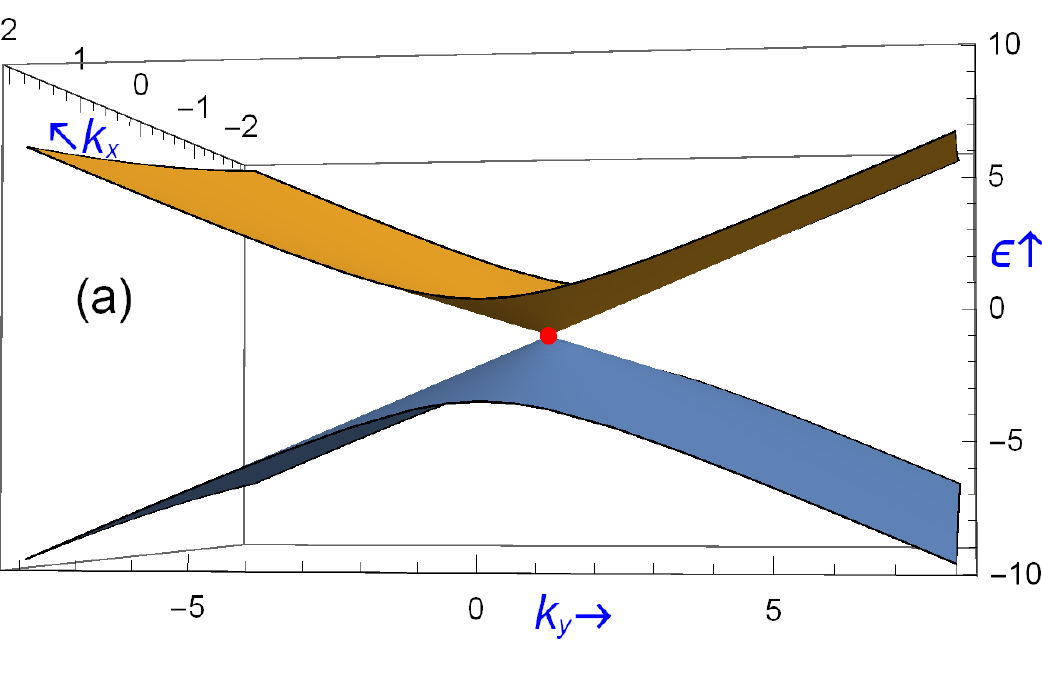}\label{FigBulkp}}\hspace{0.5cm}
\subfigure{\includegraphics[width=3.1in]{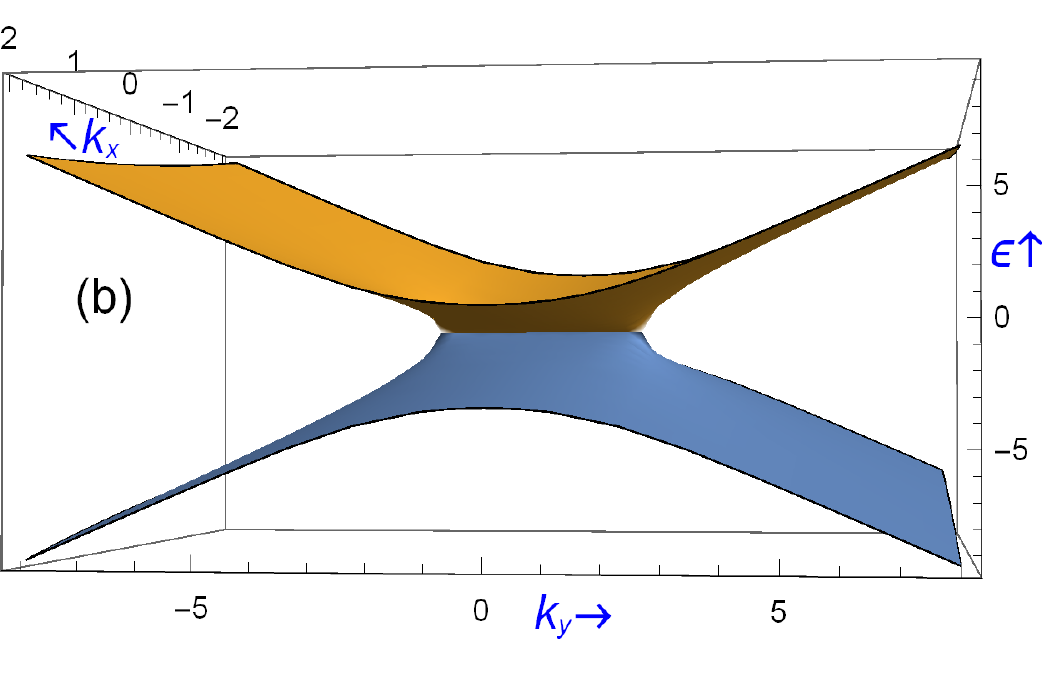}\label{FigBulka}}
\caption{Quasiparticle energy dispersion for Dirac fermions. (a)
With disorder that satisfy \Eq{Eqcondc}; (b) With disorder that
violate \Eq{Eqcondc}. Here, the momentum and energy are
dimensionless. $w = 1/\sqrt{2}$ and $\eta_- = 2$.} \label{FigBulk}
\end{figure*}

It is difficult to obtain exact analytical solutions of
$\Sigma_{11}\xk{\ep}$ and $\Sigma_{22}\xk{\ep}$. We now take the
zero-energy limit, and find that the two components of fermion
self-energy are given by
\begin{widetext}
\bea \Sigma_{11}(\ep=0) &=& \mp 2i\La\sqrt{1-w^2}\zk{\beta_+
\xk{1-w}+\xk{\beta_x+\beta_y}\xk{1+w}}\frac{1}{h}
\exp\zk{-2\xk{1-w^2}/h}\equiv \mp i\Ga_+
\label{sg11ss}, \\
\Sigma_{22}(\ep=0)&=&\mp 2i\La \sqrt{1-w^2} \zk{\beta_-
\xk{1+w}+\xk{\beta_x+\beta_y}\xk{1-w}}\frac{1}{h}
\exp\zk{-2\xk{1-w^2}/h}\equiv \mp i\Ga_- \label{sg22ss}, \eea where
\bea h = \beta_+\xk{1-w}^2+\beta_-\xk{1+w}^2 +
2\xk{\beta_x+\beta_y}\xk{1-w^2}. \eea \end{widetext} Therefore, the
retarded self-energy takes the form \bea \Sigma^{\rm
R}\xk{\ep=0}=\left[
                       \begin{array}{cc}
                         -i\Ga_+ & 0 \\
                         0 & -i\Ga_- \\
                       \end{array}
                     \right],
\eea where $\Ga_+ $ and $\Ga_-$ are two constant energy scales
defined by \Eq{sg11ss} and \Eq{sg22ss}, respectively. It is trivial
to check that \Eq{EqrelaS} is satisfied by \Eq{sg11ss} and
\Eq{sg22ss} and, therefore, by $\Ga_+$ and $\Ga_-$. For
illustration, we rewrite \Eq{EqrelaS} explicitly in terms of $\Ga_+$
and $\Ga_-$: \bea
\xk{\frac{\beta_x+\beta_y}{1+w}+\frac{\beta_-}{1-w}}\Ga_+ =
\xk{\frac{\beta_x+\beta_y}{1-w} + \frac{\beta_+}{1+w}}
\Ga_-\label{Eqrela}. \eea This result tells us immediately that
$\Ga_+=\Ga_-$ when the disorder parameters satisfy
\bea\beta_-\xk{1+w} - \beta_+\xk{1-w} = 2w\xk{\beta_x+\beta_y}.
\label{Eqcondc} \eea In this case, the Dirac point is robust, and
the strongly disordered phase can be identified as a well-defined
CDM \cite{Fradkin1, Fradkin2, shindou-murakami, ominato-koshino,
goswami-chakravarty, herbut-disorder, brouwer, Syzranov15A, Roy14,
Pixley15, Syzranov15B}. Once \Eq{Eqcondc} is violated, we always
have $\Ga_+ \neq \Ga_-$. This indicates that electrons in two
orbitals have different scattering rates, similar to the case only
the $\s_z=+1$ orbital is disordered.

\section{Fermi arc versus Dirac point \label{Sec:Farc}}

In this section, we discuss how the Fermi surface is influenced if
the condition \Eq{Eqcondc} is satisfied and violated.

If the condition \Eq{Eqcondc} is violated, we know that $\Ga_+ \neq
\Ga_-$. Adding the self-energy to the free Dirac Hamiltonian, we
obtain the total Hamiltonian $H\xk{\v p,\om}=H_0\xk{\v
p}+\Sigma\xk{\om}$. At zero energy, $\om=0$, the Hamiltonian \bea
H\xk{\v p} = \xk{wv_x p_x-i\eta_+}\s_0 + \xk{v_x p_x -
i\eta_-}\sigma_z + v_y p_y \sigma_x, \nn \\ \label{EqtotalH} \eea
where $\eta_{\pm} = \xk{\Ga_+\pm \Ga_-}/2$. Solving the equation
$\det\zk{{\cal E'}-H\xk{\v p}}=0$ leads us to the following
quasiparticle energy dispersion \bea {\cal E'}_{\pm}\xk{\v
p}=\xk{wv_x p_x-i\eta_+} \pm \sqrt{\xk{v_x p_x-i\eta_-}^2+v_y^2
p_y^2}.\label{EqComdisp} \eea Since $\Ga_+ \neq \Ga_-$, both
$\eta_+$ and $\eta_-$ are nonzero. As a result, ${\cal
E'}_{\pm}\xk{\v p}$ now has a complex value. The genuine
quasiparticle energy-momentum relation \cite{ShenNonH, KoziiNonH,
PapajNonH} corresponds to its real part $\Re\zk{{\cal
E'}_{\pm}\xk{\v p}}$, which is found to have the form \bea {\cal
E'}_{\pm}^{\rm R}=w v_x p_x \pm \frac{1}{\sqrt{2}}\sqrt{\sqrt{{\cal
E}_0^2 + 4\eta_-^2 v_x^2 p_x^2} + {\cal E}_0 }, \label{EqComdispRP}
\eea where ${\cal E}_0=v_x^2 p_x^2+v_y^2 p_y^2-\eta_-^2$. This
energy-momentum relation is plotted in \Fig{FigBulka}, and for
comparison we also plot the energy dispersion of \Eq{Eqdisp} in
\Fig{FigBulkp}. It is clear that the Dirac point located at $p_x =
p_y = 0$ is converted by disorder into a line. Setting the
quasiparticle energies in the conduction and valence bands to be
equal, we find that this line is described by \bea p_x=0,\quad -
\frac{\abs{\eta_-}}{v_y}\leq p_y \leq \frac{\abs{\eta_-}}{v_y}. \eea
Two bands are degenerate along this line, which is the Fermi arc
\cite{ShenNonH, KoziiNonH, PapajNonH}. The two end points of this
arc are $\xk{k_x,k_y}=\xk{0, \pm \abs{\eta_-}/v_y}$. These two
points, commonly identified as exceptional points, can be regarded
as the result of Dirac point splitting \cite{Heiss2012, ShenNonHbt}.
Beyond these two points, ${\cal E'}_{+}^{\rm R}>{\cal E'}_{-}^{\rm
R}$, and a band gap ${\cal E'}_g={\cal E'}_{+}^{\rm R}-{\cal
E'}_{-}^{\rm R}$ is generated, such that \bea {\cal
E'}_g=\sqrt{2}\sqrt{ \sqrt{{\cal E}_0^2+4\eta_-^2v_x^2 p_x^2}+{\cal
E}_0 }.\label{Eqgap} \eea The increase of this gap with momentum can
be clearly seen in \Fig{FigBulkp}.

From the above analysis, we know that a bulk Fermi arc appears
whenever two orbitals acquire different disorder scattering rates.
However, once the condition \Eq{Eqcondc} is met, two distinct
orbitals have exactly the same disorder scattering rate, i.e..
$\Ga_-=\Ga_+=\Ga_0$. As a result, the energy dispersion is the
direct sum of tilted Dirac fermion energy and a constant damping
rate, namely \bea {\cal E'}_{\pm} &=& w v_{x}p_{x}\pm
\sqrt{v_{x}^{2}p_{x}^{2} + v_{y}^{2}p_{y}^{2}}-i\Ga_0 \nn \\
&=& {\cal E}_{\pm} -i\Ga_0. \eea In this case, there is no Fermi
arc. This feature occurs in the strongly disordered phase of
conventional Dirac fermion system \cite{Fradkin1, Fradkin2,
shindou-murakami, ominato-koshino, goswami-chakravarty,
herbut-disorder, brouwer, Syzranov15A, Roy14, Pixley15,
Syzranov15B}, in which the symmetry between two orbitals ensures
that the two scattering rates are equal. In the model considered in
this paper, the two orbitals of tilted Dirac fermion are not related
by any symmetry and physically distinct. It is the coexistence of
different types of disorder that generates the same scattering rate
for these two distinct orbitals.

We regard the condition \Eq{Eqcondc} as a criterion for judging the
presence or absence of bulk Fermi arc in tilted Dirac fermion system
supporting independent orbitals. The system exhibits isolated Dirac
point if \Eq{Eqcondc} is satisfied, and bulk Fermi arc otherwise.
This is the main result of our paper. The condition \Eq{Eqcondc}
corresponds to a four-dimensional subspace of a five-dimensional
space spanned by the parameters $\beta_x$, $\beta_y$, $\beta_{+}$,
$\beta_{-}$, and $w$. For a given sample, these five parameters take
certain constant values. The Dirac point is stable only when these
parameters lie in this subspace. In this respect, the appearance of a
bulk Fermi arc is quite generic.

\section{Summary and discussion \label{Sec:Summ}}

In summary, we have presented a RG analysis of four types of
disorder allowed to exist by itself in 2D tilted Dirac fermion
systems. Our results indicate that, when only one type of disorder
exists, intraorbital disorder scattering can produce a bulk Fermi
arc, consistent with previous work of Ref.~\cite{PapajNonH}. Such an
arc is generated as long as the Dirac fermions from two orbitals are
not related by any symmetry. By contrast, in the case of
interorbital disorder scattering, the Dirac cone remains intact, and
no Fermi arc appears. Instead, interorbital disorder leads to
logarithmic or power-law quantum corrections to such quantities as
DOS and specific heat of tilted Dirac fermions. We have also
examined the mutual influence of different disorders, and showed
that the coexistence of two or more types of disorder dynamically
generate the rest types of disorder. Consequently, unless there is
strictly one single type of disorder, the system inevitably contains
all the four types of disorder. The interplay of different disorders
bring about physics not reported previously. Interestingly, the
Fermi arc does not always replace the Dirac point, even if the
fermions of distinct orbitals are unrelated. We obtain a condition
for the emergence of a bulk Fermi arc. The tilted Dirac fermion system
exhibits isolated Dirac point if this condition is satisfied, and a
bulk Fermi arc is formed when this condition is violated.

We then remark on possible future research projects. The formation
of a bulk Fermi arc due to different disorder scattering rates has
been studied by several groups. The topological band theory of
different scattering rates is put forward in Ref.~\cite{ShenNonHbt}.
The authors of Ref.~\cite{KoziiNonH} considered the possibility of
inducing a Fermi arc by the electron-phonon interaction in 2D untilted
Dirac fermion system. The disorder induced Fermi arc is predicted to
emerge in 3D tilted Weyl fermion systems \cite{Zyuzin18} and a 2D
tilted Dirac-fermion system with only the $\s_z=-1$ orbital being
disordered. Moreover, there is experimental evidence of a bulk Fermi
arc produced by the splitting of one single Dirac point into a pair
of exceptional points \cite{hyZhou}. So far, it remains unclear
whether such a Fermi arc can be realized in other Dirac/Weyl SMs,
such as semi-Dirac SM \cite{Dietl2008PRL, Banerjee2009PRL,
Montambaux2009EPJB, Montambaux2009PRB, Pardo2009PRL}. This problem
deserves further investigation.

Finally, we would like to address another interesting question: How
should we identify the strongly disordered phase that features two
different fermion scattering rates? In previous studies, a strongly
disordered phase with one universal scattering rate is usually
identified as a CDM phase \cite{Fradkin1, Fradkin2,
shindou-murakami, ominato-koshino, goswami-chakravarty,
herbut-disorder, brouwer, Syzranov15A, Roy14, Pixley15,
Syzranov15B}. There is no bulk Fermi arc in such a phase. For the
Dirac fermion system considered in this paper, the strongly
disordered phase cannot be regarded as CDM, although a finite
zero-energy DOS is generated. The appearance of two distinct
disorder scattering rates produces a Fermi arc, which makes such a
strongly disordered phase quite different from the conventional one
\cite{Fradkin1, Fradkin2, shindou-murakami, ominato-koshino,
goswami-chakravarty, herbut-disorder, brouwer, Syzranov15A, Roy14,
Pixley15, Syzranov15B}.

\section{Acknowledgments}

We would like to thank Jing-Rong Wang for helpful discussion, and
acknowledge the support provided by the National Natural Science
Foundation of China under Grants 11375168 and 11574285. G.-Z.L. is
partly supported by the Fundamental Research Funds for the Central
Universities (P. R. China) under Grant WK2030040085.

\newpage

\appendix

\begin{widetext}

\section{Derivation of coupled RG equations \label{Sec:ApenRG}}

We first discuss how to determine the four types of bilinear
fermion-disorder coupling, and then derive the RG equations given by
\Eqs{RGvx}-(\ref{RGbetay}) by calculating the diagrams of
Fig.~\ref{FigAll}(a)-Fig.~\ref{FigAll}(d).

The calculation will be based on the free fermion propagator of
fermions: \bea G_{0}\xk{ip_0,\v p} &=& \frac{1}{-ip_0\sigma_0 +
(\sigma_z + w\sigma_0)v_x p_x + v_y \sigma_x p_y } \nonumber \\
&=& \frac{\xk{ip_0 - wv_x p_x}\sigma_0+v_x p_x\s_z+ v_y
p_y\sigma_x}{\xk{p_0+iwv_x p_x}^2 +v_x^2 p_x^2+ v_y^2  p_y^2}.
\label{Eqfreepro} \eea

\subsection{Determining the matrices for fermion-disorder coupling \label{Sec:dismatr}}

Intraorbital disorder can be formally defined via the matrix $\ga =
A\s_0 + B\s_z$, where $A$ and $B$ are both real to ensure that $\ga$
is hermitian. Now, we consider the loop correction of
Fig.~\ref{FigAll}(b) with disorder vertex $\ga$, and find that \bea
\delta\Delta_{\ga}^{(b)} &=& \Delta_{\ga}^2
\int'\frac{d^2\mathbf{p}}{(2\pi)^2} \ga
G_0(0,\mathbf{p})\ga G_0(0,\mathbf{p})\ga \nn \\
&=& \frac{\Delta_{\ga}^2}{2\pi v_yv_x\xk{1-w^2}^{3/2}}\left(
                                                  \begin{array}{cc}
                                                  \xk{1-w}\xk{A-Bw}\xk{A+B}^2 & 0 \\
                                                  0 & \xk{1+w}\xk{A-Bw}\xk{A-B}^2 \\
                                                  \end{array}
                                                   \right),\label{Eqdism}
\eea where \bea \int'd^{2}\v p = \int_{\La e^{-\ell}}^{\La}\abs{\v
p}d \abs{\v p}\int_{0}^{2\pi}d\th, \quad \xk{p_x=\abs{\v p}\cos\th,
p_y=\abs{\v p}\sin\th}. \eea To make RG analysis self-consistent,
the matrix appearing in \Eq{Eqdism} must be proportional to
$\ga=\left(
            \begin{array}{cc}
                  A+B & 0 \\
                 0& A-B \\
        \end{array}
      \right)$
for any $w\in \xk{-1,1}$. It is easy to verify that this condition
is fulfilled only when $A = \pm B$. Therefore, we should choose the
matrices as $\ga = A\xk{\s_0+\s_z}\equiv \s_+$ and $\ga =A \xk{\s_0
- \s_z}\equiv \s_-$. The constant $A$ can be absorbed into the
effective disorder parameter defined by \Eq{Eqeffcoup}. As a result,
the value of $A$ does not affect the RG equations, and we simply set
$A = 1/2$.

For interorbital disorder denoted by $\ga = A'\s_x + B'\s_y$,
similar analysis indicates that self-closed RG analysis is obtained
only when $A'=0$ or $B'=0$. This implies that $\ga=\s_x$ or
$\ga=\s_y$.

\subsection{Interaction corrections and RG equations \label{Sec:RGeqs}}

After specifying the disorder vertices, we proceed to calculate the
interaction corrections and to derive the coupled RG equations.

Fig.~\ref{FigAll}(a) is the fermion self-energy correction caused by
disorder scattering. It can be written as \bea
\Sigma_{\mathrm{dis}}(i p_0) &=& -\sum_i\Delta_{i}
\int'\frac{d^{2}\mathbf{p}}{(2\pi)^{2}}\s_i G_{0}\xk{ip_0,\v p}\s_i
\nn\\ &=& \frac{-i p_0 \ell}{2\pi v_y v_x\sqrt{1-w^2}}
\zk{\xk{\frac{\De_+}{1+w}+\frac{\De_x+\De_y}{1-w}}\s_+
+\xk{\frac{\De_-}{1-w}+\frac{\De_x+\De_y}{1+w}}\s_-}.\label{Eqseenone}
\eea Fig.~\ref{FigAll}(b)-Fig.~\ref{FigAll}(d) represent the
corrections to fermion-disorder vertices. We find that \bea
\delta\Delta_i^{(b)} &=& \Delta_i\xk{\psi_m^\dagger\s_i\psi_m}
\psi_n^\dagger\sum_j \Delta_j \int'\frac{d^2\mathbf{p}}{(2\pi)^2}
\s_j G_0(0,\mathbf{p})\s_i G_0(0,\mathbf{p})\s_j \psi_n. \eea For
given $\s_i$, the corrections to $\Delta_i$ are given by \bea
\delta\Delta_i^{(b)}= \left\{
\begin{array}{ll}
0 & \hbox{$\s_i=\s_x$,} \\*[0.4cm]
\dfrac{\Delta_y\xk{\Delta_x-\Delta_y}\ell}{2\pi
v_xv_y\sqrt{1-w^2}}\xk{\psi_m^\dagger\s_y\psi_m}\xk{\psi_n^\dagger\s_y\psi_n}
& \hbox{$\s_i=\s_y$,} \\*[0.4cm] \dfrac{\Delta_+\ell}{4\pi
v_xv_y\sqrt{1-w^2}}\Bigg\{\zk{\De_x+\De_y+\De_+\xk{\dfrac{1-w}{1+w}}
}\xk{\psi_m^\dagger\s_+\psi_m}\xk{\psi_n^\dagger\s_+\psi_n}
\\*[0.4cm]+\zk{\xk{\De_x+\De_y}\xk{\dfrac{1-w}{1+w}} + \De_-}
\xk{\psi_m^\dagger\s_+\psi_m}\xk{\psi_n^\dagger\s_-\psi_n}\Bigg\}
& \hbox{$\s_i=\s_+$,}\\*[0.4cm] \dfrac{\Delta_-\ell}{4\pi
v_xv_y\sqrt{1-w^2}}\Bigg\{\zk{\De_++\xk{\De_x+\De_y}\xk{\dfrac{1+w}{1-w}}
}\xk{\psi_m^\dagger\s_+\psi_m}\xk{\psi_n^\dagger\s_-\psi_n}
\\*[0.4cm] +\zk{\De_x+\De_y+\De_-\xk{
\dfrac{1+w}{1-w}}}\xk{\psi_m^\dagger\s_-\psi_m}\xk{\psi_n^\dagger\s_-\psi_n}
\Bigg\} & \hbox{$\s_i=\s_-$.}
\end{array}
\right. \eea Here, notice that the terms proportional to
$\xk{\psi_m^\dagger\s_+\psi_m}\xk{\psi_n^\dagger\s_- \psi_n}$ are
forbidden within the replica formalism. We simply discard these
terms hereafter.

For the diagrams of Fig.~\ref{FigAll}(c)+Fig.~\ref{FigAll}(d), we
get \bea \delta\Delta^{(c)+(d)} &=& \sum_{ij}\Delta_i
\Delta_j\int\frac{d^{2}\mathbf{p}}{(2\pi)^{2}}\psi^{\dagger}_m\l[\s_i
G_0(0,\mathbf{p})\s_j\r]\psi_m\psi^{\dagger}_n\l[\s_j
G_0(0,\mathbf{p})\s_i + \s_i G_0(0,-\mathbf{p})\s_j\r]
\psi_n\label{Eqc+d}, \eea One can see that
$\delta\Delta^{(c)+(d)}=0$ for $\s_i=\s_j$ because of the relation
$G_0(0,-\mathbf{p}) = -G_0(0,\mathbf{p})$. Thus we only need to
consider $\s_i\neq\s_j$, which contains six pairs as
$\l(\s_i,\s_j\r)= \{\xk{\s_x,\s_y}, \xk{\s_x,\s_+}, \xk{\s_x,\s_-},
\xk{\s_y,\s_+}, \xk{\s_y,\s_-}, \xk{\s_+,\s_-}\}$. After computing
\Eq{Eqc+d} for all of these pairs, we eventually obtain \bea
\delta\Delta^{(c)+(d)}= \left\{
\begin{array}{ll}
\dfrac{\Delta_x \Delta_y\ell}{2\pi v_x
v_y\sqrt{1-w^2}}\xk{\dfrac{1+w}{1-w}}
\xk{\psi^{\dagger}_m\s_+\psi_m} \xk{\psi^{\dagger}_n\s_+\psi_n} +
\\*[0.4cm] \dfrac{\Delta_x \Delta_y\ell}{2\pi
v_xv_y\sqrt{1-w^2}}\xk{\dfrac{1-w}{1+w}}
\xk{\psi^{\dagger}_m\s_-\psi_m} \xk{\psi^{\dagger}_n\s_-\psi_n} &
\hbox{$\l(\s_i,\s_j\r)= \xk{\s_x,\s_y}$,}
\\*[0.6cm]
\dfrac{\Delta_x \Delta_+}{8\pi v_x v_y\sqrt{1-w^2}}
\xk{\dfrac{1-w}{1+w}}\xk{\psi^{\dagger}_m\s_y\psi_m}
\xk{\psi^{\dagger}_n\s_y\psi_n} & \hbox{$\l(\s_i,\s_j\r)=
\xk{\s_x,\s_+}$,}
\\*[0.6cm]
\dfrac{\Delta_x \Delta_- }{8\pi
v_xv_y\sqrt{1-w^2}}\xk{\dfrac{1+w}{1-w}}\xk{\psi^{\dagger}_m\s_y\psi_m}
\xk{\psi^{\dagger}_n\s_y\psi_n} & \hbox{$\l(\s_i,\s_j\r)=
\xk{\s_x,\s_-}$,}
\\*[0.6cm]
\dfrac{\Delta_y \Delta_+ \ell}{8\pi v_xv_y\sqrt{1-w^2}}\xk{\dfrac{1-w}{1+w}}
\xk{\psi^{\dagger}_m\s_x\psi_m}\xk{\psi^{\dagger}_n\s_x\psi_n}+ \\*[0.4cm]
\dfrac{\Delta_y \Delta_+ \ell}{2\pi v_xv_y\sqrt{1-w^2}}
\xk{\psi^{\dagger}_m\s_+\psi_m}\xk{\psi^{\dagger}_n\s_+\psi_n}
& \hbox{$\l(\s_i,\s_j\r)= \xk{\s_y,\s_+}$,}
\\*[0.6cm]
\dfrac{\Delta_y \Delta_- \ell}{8\pi v_xv_y\sqrt{1-w^2}}\xk{\dfrac{1+w}{1-w}}
\xk{\psi^{\dagger}_m\s_x\psi_m}\xk{\psi^{\dagger}_n\s_x\psi_n}+ \\*[0.4cm]
\dfrac{\Delta_y \Delta_- \ell}{2\pi v_xv_y\sqrt{1-w^2}}
\xk{\psi^{\dagger}_m\s_-\psi_m}\xk{\psi^{\dagger}_n\s_-\psi_n}
& \hbox{$\l(\s_i,\s_j\r)= \xk{\s_y,\s_-}$,}
\\*[0.6cm]
\dfrac{\Delta_+ \Delta_- \ell}{8\pi v_xv_y\sqrt{1-w^2}}
\xk{\psi^{\dagger}_m\s_y\psi_m}\xk{\psi^{\dagger}_n\s_y\psi_n} &
\hbox{$\l(\s_i,\s_j\r)= \xk{\s_+,\s_-}$.}
\end{array}
\right. \eea Before going further, we re-define the disorder
strength parameters: \bea \beta_{i} \equiv \frac{\Delta_{i}}{2\pi
v_xv_y\sqrt{1-w^2}}. \eea

We insert all the one-loop corrections into the free action. After
making the transformations \bea
&&\tilde{\om}=e^{-\ell}\om,\,\,\tilde{k}_i=e^{-\ell}k_i,\,\,
\tilde{\psi}_1\xk{\tilde{p}}=\sqrt{Z_1}\psi_1\xk{p},\,\,
\tilde{\psi}_2\xk{\tilde{p}}=\sqrt{Z_{2}}\psi_2\xk{p},\,\,
v_x(1+w)=Z_{w^+}\tilde{v}_x(1+\tilde{w}) \nn \\ &&
v_x(1-w)=Z_{w^-}\tilde{v}_x(1+\tilde{w}),\,\,\tilde{v}_{y}=Z_{v_{y}}v_y,\,\,
\tilde{\Delta}_i=Z_{\Delta_i}\Delta,\,\,
\tilde{\beta}_i=\frac{Z_{\Delta_i}}{Z_{v_{y}}\sqrt{Z_{w^+}Z_{w^-}}}
\beta_i = Z_{\beta_i}\beta_i, \eea we can recast the renormalized
action in its original form. Based on these manipulations, we derive
the flow equations: \bea
\frac{d\xk{wv_x}}{d\ell}&=&-\zk{\frac{\beta_+-\beta_-}{2}
-\frac{\beta_x+\beta_y}{2}\xk{\frac{1-w}{1+w}-\frac{1+w}{1-w}} }v_x
\label{RGtap}, \\
\frac{dv_x}{d\ell}&=&-\zk{\frac{\beta_++\beta_-}{2}
+\frac{\beta_x+\beta_y}{2}\xk{\frac{1+w}{1-w}+\frac{1-w}{1+w}}}
v_x\label{RGvxap},
\\
\frac{d v_{y}}{d\ell} &=& -\frac{1}{2} \zk{\frac{\xk{\beta_{x} +
\beta_{y}}}{1-w}+\frac{\beta_{+}}{1+w} +
\frac{\xk{\beta_{x}+\beta_{y}}}{1+w}+\frac{\beta_{-}}{1-w}}v_{y}
\label{RGvyap}, \\
\frac{d \beta_x}{d\ell}&=& \dfrac{\beta_y}{2}\zk{\beta_+
\xk{\frac{1-w}{1+w}} +\beta_-\xk{\frac{1+w}{1-w}}},
\label{RGbetaxap} \\
\frac{d \beta_y}{d\ell} &=& -2\xk{\beta_y-\beta_x}\beta_y +
\dfrac{\beta_x}{2} \zk{\beta_+ \xk{\frac{1-w}{1+w}}
+\beta_-\xk{\frac{1+w}{1-w}}} + \dfrac{\beta_+ \beta_-}{2},
\label{RGbetayap}
\\
\frac{d \beta_+}{d\ell}&=&\frac{-\beta _+^2 w}{1+w}+\beta _+
\left[\frac{\beta _- }{1-w}-\frac{2 w \left(\beta _x+\beta
_y\right)}{1-w^2}+\beta _x+3 \beta _y\right]+\frac{2 \beta _x \beta
_y\left(1+w\right) }{1-w},\label{RGbeta+ap}
\\
\frac{d \beta_-}{d\ell}&=&\frac{\beta _-^2 w}{1-w}+\beta _-
\left[\frac{\beta _+ }{1+w}+\frac{2 w \left(\beta _x+\beta
_y\right)}{1-w^2}+\beta _x+3 \beta _y\right]+\frac{2 \beta _x \beta
_y\left(1-w\right) }{1+w} .\label{RGbeta-ap} \eea
\Eqs{RGvxap}~-~(\ref{RGbetayap}) are just
\Eqs{RGvx}~-~(\ref{RGbetay}) presented in the main text. Combining
\Eq{RGtap} and \Eq{RGvxap} leads to \Eq{RGw} of the main text.
Combining \Eq{RGbeta+ap} and \Eq{RGbeta-ap} leads to \Eq{RGbetapm}
of the main text.
\end{widetext}

\section{Proof of the nonrenormalization of RGP \label{Sec:ApenRGP}}

According to \Eq{RGbetaxs}, the disorder parameter of RGP is not
renormalized and remains a marginal perturbation at one-loop level.
The aim of this Appendix is to illustrate that this conclusion is
valid at any order of loop expansion. To prove this, it is more
convenient to first write the action in the imaginary-time
formalism: \bea S&=&\int d\tau d^2\v x \psi^\dagger(\v x)
\zk{\s_0\pa_{\tau}-i(\sigma_z +w \sigma_0)v_x
\partial_x \r.\nn\\&&\l.-iv_y \sigma_x\partial_y -v_d A(\v x)\s_x }
\psi (\v x). \eea Be using the replica method, we find that $v_d$ is
connected to the effective disorder coupling via the relation \bea
\beta_x=\frac{v_d^2}{2\pi v_y v_x \sqrt{1-w^2}}. \eea Discarding the
energy-independent constants, we only need to verify that \bea
\frac{d}{d\ell}\xk{\frac{v_d^2}{v_y v_x \sqrt{1-w^2}}}=0. \eea
Making the local gauge transformation \bea \psi_i \rightarrow \psi_i
e^{i\xi(y)}, \quad A(\v x)\rightarrow A(\v x)+ \pa_y \xi(y), \eea we
re-express the action as \bea S &=& \int d\tau d^2\v x\dk{
\psi^\dagger(\v x)\zk{\s_0\pa_{\tau}-i(\sigma_z + w \sigma_0)v_x
\partial_x \r.\r.\nn\\&&\l.\l.-iv_y \sigma_x\partial_y -v_d A(\v
x)\s_x}\psi(\v x)+\psi^\dagger(\v x)\s_x \psi(\v x)\r.
\nn\\&&\times\l.\xk{v_y-v_d}\pa_y\xi(y)}. \eea To preserve the gauge
invariance, one must demand that \bea v_y\xk{\ell} = v_d\xk{\ell}.
\eea The validity of this identify is loop independent. It is now
obvious that \bea \frac{d v_y}{d\ell}=\frac{d v_d}{d\ell}.
\label{gaugeeq} \eea

In addition, a crucial character of static disorder is that the
fermion self-energy is independent of momenta. Therefore, at any
order of loop expansion, integrating out the fast modes defined
within the shell $\xk{\La e^{-\ell},\La}$ leads to an effective,
renormalized action
\begin{widetext}
\bea
S &=&\int d\tau d^2\v x \psi^\dagger(\v x) \zk{ \left(
                                              \begin{array}{cc}
                                               1+\Sigma_{1}(\ell) & 0 \\
                                               0 & 1+\Sigma_{2}(\ell) \\
                                                  \end{array}
                                             \right)
\pa_{\tau}-i(\sigma_z +w \sigma_0)v_x\partial_x-iv_y
\sigma_x\partial_y +v_d\xk{1+\de v_d} A(\v x)\s_x}\psi(\v x) \nn
\\ &=& \int d\tau d^2\v x \Big\{\psi_1^\dagger(\v x)
\zk{\xk{1+\Sigma_{1}(\ell)}\pa_{\tau}-iv_x(1 + w)\partial_x}\psi_1
(\v x)+\psi_2^\dagger(\v x)\zk{\xk{1+\Sigma_{2}(\ell)}
\pa_{\tau}+iv_x(1 - w)\partial_x}\psi_2 (\v x) \nn \\ && -iv_y\zk{
\psi_1^\dagger(\v x)\partial_y \psi_2(\v x)+\psi_2^\dagger(\v x)
\partial_y \psi_1(\v x)} +v_d\xk{1+\de v_d} A(\v
x) \zk{\psi_1^\dagger(\v x) \psi_2(\v x)+\psi_2^\dagger(\v x)
\psi_1(\v x)}\Big\}. \eea\end{widetext} We re-scale the coordinates
and field operators as follows: $\tau \rightarrow e^{\ell} \tau,
\quad \v x \rightarrow e^{\ell} \v x, \quad \psi_i\rightarrow
Z_i^{-1/2}\psi_i,\quad v_x\xk{1-w} \rightarrow
Z_{w^-}^{-1}v_x\xk{1-w},\quad v_x\xk{1+w} \rightarrow
Z_{w^+}^{-1}v_x\xk{1+w},\quad v_y \rightarrow Z_{v_y}^{-1}v_y$.
Straightforward calculations give rise to \bea Z_{w^+} =
\frac{1}{1+\Sigma_{1}(\ell)}, \quad
Z_{w^-} = \frac{1}{1+\Sigma_{2}(\ell)},\nn\\
Z_{v_y} = \frac{1}{\sqrt{1+\Sigma_{1}(\ell)}
\sqrt{1+\Sigma_{2}(\ell)}}. \eea Apparently, the following identify
holds \bea Z_{v_y}^2(\ell)=Z_{w^+}(\ell)Z_{w^-}(\ell), \eea which
directly leads to \bea \frac{d\ln v_y^2}{d\ell}&=&\frac{d
Z_{v_y}^2(\ell)}{d\ell}\bigg|_{\ell=0} \nn\\&=&Z_{w^+}(\ell)\frac{d
Z_{w^-}(\ell)}{d\ell}\bigg|_{\ell=0}+Z_{w^-}(\ell)\frac{d
Z_{w^+}(\ell)}{d\ell}\bigg|_{\ell=0}\nn\\&=&\frac{d
Z_{w^+}(\ell)}{d\ell}\bigg|_{\ell=0}+\frac{d
Z_{w^-}(\ell)}{d\ell}\bigg|_{\ell=0} \nn\\ &=&\frac{d\ln v_x\xk{1-w}
}{d\ell}+\frac{d\ln v_x\xk{1+w} }{d\ell}\nn\\&=&\frac{d\ln
v_x^2\xk{1-w^2}}{d\ell}. \eea This can be further written as \bea
\frac{d\ln v_y}{d\ell}=\frac{d\ln v_x \sqrt{1-w^2}}{d\ell}.
\label{renoreq} \eea Based on \Eq{gaugeeq} and \Eq{renoreq}, we find
that \bea &&\dfrac{d}{d\ell} \ln \xk{\dfrac{v_d^2}{v_y
v_x\sqrt{1-w^2}}} = 0. \eea Thus, we conclude that \bea \dfrac{d
\beta_x}{d\ell}=0 \eea is valid at any order of loop expansion.

\section{DOS and Specific heat \label{Sec:Apenob}}

Here we calculate the fermion DOS and specific heat, and then use
the results to discuss the impact of interorbital disorder.

\subsection{Low-energy DOS \label{Sec:ApenDOS}}

The DOS $\rho(\omega)$ is defined as
\begin{eqnarray}
\rho(\omega)=-N\int \frac{d^2 \mathbf{k}}{(2 \pi)^2}
\trace{\Im{G^{R}(\omega,k_x,k_y)}},
\end{eqnarray}
where $N=2$, represents the flavor of Dirac
fermions. Carrying out analytic continuation $i p_0\rightarrow \omega +
i\gamma$ ($\gamma\rightarrow 0$) to \Eq{Eqfreepro}, we get
\begin{eqnarray}
G_0^{R}(\omega,\mathbf{k}) &=& \left[\mathcal{P}
\frac{1}{k_{x}^2v_x^2 + k_{y}^2 v_y^2 - \omega'^2}-i\pi
\sgn(\omega')\r.\nn\\ && \times \l.\delta(-\omega'^2+
k_{x}^2v_x^2+k_{y}^2v_y^2)\right]\nn\\&&\times\xk{\omega'\s_0 +
k_{x}v_x\s_z + v_yk_y\s_x},
\end{eqnarray}
where $\om'=\om-wv_xk_x$. It is now easy to get the spectral
function
\begin{eqnarray}
A_0(\omega,\mathbf{k}) &=& -\frac{1}{\pi}\trace{\Im
G_0^{R}(\omega,\mathbf{k})} \nn \\ &=& 2|\omega'|\delta(-\omega'^2+
k_{x}^2v_x^2+k_{y}^2v_y^2).
\end{eqnarray}
The fermions DOS can be computed directly, i.e.,
\begin{eqnarray}
\rho_0(\omega) &=& N\int \frac{d^2\mathbf{k}}{(2\pi)^2}
A_0(\omega,\mathbf{k}) \nn\\ &=& \frac{\abs{\om}}{\pi v_y
v_x\left(1-w^2\right){}^{3/2}}, \label{EqDOScl}
\end{eqnarray}
which means that
\begin{eqnarray}
\rho_0(E)\propto E \propto  T,
\end{eqnarray}
where $T$ is certain temperature.

The influence of disorder is usually embodied in the quantum
corrections to the fermion velocities and the tilt parameter. To
proceed, we appeal to the transformation $\om = \La e^{-\ell}$,
where $\La$ is the UV cutoff and $\om$ is certain lower energy
scale. According to \Eq{EqDOScl}, we have \bea \frac{d\ln
\rho\xk{E}}{d\ln E}&=&1-\frac{d\ln v_x\xk{E}}{d\ln E}-\frac{d\ln
v_y\xk{E}}{d\ln E} \nn \\ &&-\frac{3 d\ln \zk{1-w^2\xk{E}}}{2d\ln E}
\nn \\ &=& \!\!1\!-\!2\zk{\beta_{x}\xk{E}\!+\!\beta_{y}\xk{E}}\!-\!
\frac{\beta_{+}\xk{E}\zk{1\!-\!w\xk{E}}}{1\!+\!w\xk{E}} \nn \\ &&-
\frac{\beta_{-}\xk{E}\zk{1+w\xk{E}}}{1-w\xk{E}}. \eea

When there is only RGP, $\beta_{x}\xk{E}=\beta_{x}\xk{\La}$, so we have
\bea \frac{d\ln \rho\xk{E}}{d\ln E} &=& 1-2\beta_{x}\xk{\La}, \eea which
gives rise to \bea \rho\xk{\omega} = \rho\xk{\La}
\xk{\frac{E}{\La}}^{1-2\beta_{x}\xk{\La}}, \eea where $\rho\xk{\La}$ is a
constant DOS defined at energy scale $\La$. This result implies that
the marginal RGP leads to a power-law enhancement of low-energy DOS.

When there is only RM, the RG solution of the effective disorder
parameter is \bea \beta_y\xk{E} = \frac{\beta_y\xk{\La}}{1+2\beta_y
\xk{\La}\ln\xk{\La/E}}, \eea where $\beta_y\xk{\La}$ can be taken as
a constant. This then yields \bea \frac{d\ln \rho\xk{E}}{d\ln E }&=&
1-2\beta_{y}\xk{E} \nn \\ &=& 1-\frac{2\beta_y\xk{\La}}{1+2\beta_y
\xk{\La}\ln\xk{\La/E}}. \eea This equation has the solution \bea
\rho\xk{E}&=&\rho\xk{\La}\xk{\frac{E}{\La}}\zk{1+2\beta_y\xk{\La}\ln\xk{\La/E}}
\nn\\ &\sim& E\ln E. \eea From this solution we conclude that the
marginally irrelevant RM only gives rise to logarithmic enhancement
to the low-energy DOS.

\subsection{Specific heat}

To calculate the specific heat, we need to first compute the free
energy. After performing functional integration, we find that the
free energy has the form
\begin{eqnarray}
F_{f}(T) &=& -2NT\sum_{\omega_{n}} \int
\frac{d^2\mathbf{p}}{(2\pi)^2} \ln\abs{\Det{G^{-1}\xk{\om_n,\v p}}}
\nn \\ &=& -T\sum_{\omega_{n}}\int\frac{d^2\mathbf{p}}{(2\pi)^2}
\ln\big[\l(\omega_{n}-iwp_x\r)^2 + v_x^2p_x^2 \nn \\ && +v_y^2p_y^2
\big],
\end{eqnarray}
where $\omega_{n}=(2n+1)\pi T$ is the Matsubara imaginary frequency.
Performing frequency summation yields
\begin{eqnarray}
F_{f}(T) &=& -2NT\int\frac{d^2\mathbf{p}}{(2\pi)^2}
\Big[\ln\left(1+e^{\beta E_-}\right) \nn \\ && +\ln\left(1+e^{\beta
E_+}\right) \Big],
\end{eqnarray}
where $E_{\pm} = w v_x p_x \pm \sqrt{v_x^2p_x^2+v_y^2p_y^2}$. This
free energy is divergent. To regularize the integral, we need to
replace $F_{f}(T)$ by $F_{f}(T)-F_{f}(0)$. After doing so, we obtain
\bea F_{f}(T) &=& -2NT\int\frac{d^2\mathbf{p}}{(2\pi)^2}
\zk{\ln\left(1+e^{\beta E_-}\right) \r.\nn \\
&&\l.+\ln\left(1+e^{-\beta E_+}\right)} \nn \\ &=& -\frac{3\zeta
(3)}{\pi v_y v_x \xk{1-w^2}^{3/2}}T^3, \eea where $\zeta (x)$ is the
Riemann zeta function. The specific heat can be obtained by
\begin{eqnarray}
C_{v}(T) = -T\frac{\partial^2 F_{f}(T)}{\partial T^2} =
\frac{18\zeta (3)T^2}{\pi v_y v_x \xk{1-w^2}^{3/2}}.
\end{eqnarray}
Employing the transformation $T = T_{\La}e^{-\ell}$, where $T_{\La}$
is the temperature corresponding to the UV cutoff $\La$, we find
\bea \frac{d\ln C_v\xk{T}}{d\ln T}&=&2-\frac{d\ln v_x\xk{T}}{d\ln T}
- \frac{d\ln v_y\xk{T}}{d\ln T}\nn \\ &&-\frac{3d\ln
\zk{1-w^2\xk{T}}}{2d\ln T} \nn \\ &=& \!2\!-\!2
\zk{\beta_{x}\xk{T}\!+\!\beta_{y}\xk{T}}\!-\! \frac{\beta_{+}\xk{T}
\zk{1\!-\!w\xk{T}}}{1\!+\!w\xk{T}}\nn\\
&& -\frac{\beta_{-}\xk{T} \zk{1+w\xk{T}}}{1-w\xk{T}}. \eea Now we
can analyze the $T$-dependence of specific heat.

When there is only RGP, one can show that \bea C_v\xk{T}= C_v
\xk{T_{\La}}\xk{\frac{T}{T_{\La}}}^{2-2\beta_{x}\xk{\La}} \sim
T^{2-2\beta_{x}\xk{\La}}. \eea

When there is only RM, we get \bea C_v\xk{T} &=& C_v\xk{T_{\La}}
\xk{\frac{T}{T_{\La}}}^2 \zk{1+2\beta_y\xk{\La}\ln\xk{T_{\La}/T}}
\nn\\ &\sim& T^2\ln T. \eea Again, RGP leads to power-law
enhancement of specific heat, whereas RM results in a logarithmic
enhancement.

\section{SCBA calculation \label{Sec:ApenSCBA}}

In this appendix, we derive the SCBA equation and then get its
solution. To make a generic analysis, the derivation will be
completed in the case that all the four types of disorder coexist.
The self-consistent equation for the fermion self-energy is given by
\begin{widetext}
\bea \Sigma(\ep)\!\! &=& \!\!\sum_i\Delta_i \int''\frac{d^2\v
k}{(2\pi)^2} \s_i \frac{1}{\ep-H_0\xk{\v k} - \Sigma(\ep)} \s_i
\label{EqSCBA}
\\ \nn\!\!&=&\!\!
\int''\!\frac{d^2\v k}{(2\pi)^2}\frac{1}{f_-^2-f_+^2+q_y^2 v_y^2}\!
\left[
  \begin{array}{cc}
   \left(\Delta_x + \Delta_y\right)\xk{f_--f_+}-\De_+\xk{f_-+f_+} & 0 \\*[0.3cm]
    0 & \De_-\xk{f_--f_+}-\left(\Delta_x+\Delta_y\right)\xk{f_-+f_+} \\
  \end{array}
\right], \eea where \bea \int''d^{2}\v k = \int_{0}^{\La}\abs{\v k}d
\abs{\v k}\int_{0}^{2\pi}d\th,\quad f_-=v_x
k_x+\dfrac{\Sigma_{11}(\ep)-\Sigma_{22}(\ep)}{2},\quad f_+=\ep-wv_x
k_x-\dfrac{\Sigma_{11}(\ep)+\Sigma_{22}(\ep)}{2}.\label{Eqexplain}
\eea \Eq{EqSCBA} is decomposed into two coupled equations:\bea
\Sigma_{11}(\ep) &=& \int''\frac{d^2\v k}{(2\pi)^2}
\frac{1}{f_-^2-f_+^2+q_y^2 v_y^2}\zk{-\De_+\xk{f_-+f_+} +
\left(\Delta_x + \Delta_y\right)\xk{f_--f_+}},
\label{EqSCBA11} \\
\Sigma_{22}(\ep) &=& \int''\frac{d^2\v k}{(2\pi)^2} \frac{1}{f_-^2 -
f_+^2 + q_y^2 v_y^2}\zk{\De_-\xk{f_--f_+}-\left(\Delta_x +
\Delta_y\right)\xk{f_-+f_+}}. \label{EqSCBA22} \eea \end{widetext}
Integrating over $\mathbf{k}$ in Eq.(D3) yields \bea g = \pm 2i\La
\sqrt{1-w^2}
\exp\zk{\frac{-2\Sigma_{11}(\ep)}{g\xk{\frac{\beta_x+\beta_y}{1-w}+
\frac{\beta_+}{1+w}}}}, \label{EqCoSCBA11} \eea where
$g=\left(1-w\right)\Sigma_{11}(\ep) +
\left(1+w\right)\Sigma_{22}(\ep)-2\ep$ and $+,-$ represent advanced
and retarded self-energy functions respectively. After completing
the integration of \Eq{EqSCBA22}, we obtain \bea g=\pm 2i\La
\sqrt{1-w^2} \exp\zk{\frac{-2\Sigma_{22}(\ep)}{g\xk{\frac{\beta_x +
\beta_y}{1+w} + \frac{\beta_-}{1-w}}}}. \label{EqCoSCBA22} \eea It
seems difficult to obtain the analytical solutions for
$\Sigma_{11}(\ep)$ and $\Sigma_{22}(\ep)$. However, we observe from
\Eq{EqCoSCBA11} and \Eq{EqCoSCBA22} that \bea &&\xk{\frac{\beta_x+\beta_y}{1+w} +
\frac{\beta_-}{1-w}}\Sigma_{11}(\ep) \nn \\ &=&
\xk{\frac{\beta_x+\beta_y}{1-w}+ \frac{\beta_+}{1+w}}
\Sigma_{22}(\ep). \label{EqrelaA} \eea If only the $\s_z=-1$ orbital
is disordered, this relation is reduced to a trivial identity $0=0$,
which gives us no new information. However, when four types of
disorder coexist, this relation provides a strong connection for the
self-energy functions of two distinct orbitals. The physical
implication of this constraint is discussed in greater detail in the
main text.

Next, we consider the solution for zero energy, and we focus on the
retarded self-energy. Substitute \Eq{EqrelaA} to \Eq{EqCoSCBA11} and
\Eq{EqCoSCBA22} with $\ep=0$ we get \bea \Sigma_{11}^R(\ep=0)&=&
-2i\La\sqrt{1-w^2}\zk{\beta_+\xk{1-w}+
\xk{\beta_x+\beta_y}\r.\nn\\&&\times\l.\xk{1+w}} \frac{1}{h}
\exp\zk{-2\xk{1-w^2}/h }
\label{sg11ssA}, \\
\Sigma_{22}^R(\ep=0)&=&
-2i\La\sqrt{1-w^2}\zk{\beta_-\xk{1+w}+\xk{\beta_x+\beta_y}\r.\nn\\&&\times\l.\xk{1-w}
} \frac{1}{h} \exp\zk{-2\xk{1-w^2}/h },\label{sg22ssA} \eea where
\bea
h&=&\beta_+\xk{1-w}^2+\beta_-\xk{1+w}^2\nn\\&&+2\xk{\beta_x+\beta_y}\xk{1-w^2}.
\eea Therefore, \bea \Sigma^{R}\xk{\ep=0}=\left[
                       \begin{array}{cc}
                         -i\Ga_+ & 0 \\
                         0 & -i\Ga_- \\
                       \end{array}
                     \right],
\eea where $\Ga_+$ and $\Ga_-$ correspond to two energy scales that
can be obtained from \Eq{sg11ssA} and \Eq{sg22ssA}. In the special
case that only the $\s_z=-1$ orbital is disordered, one can easily
get \bea \Ga_+=0,\quad \Ga_-=2\La\sqrt{\frac{1-w}{1+w}}\exp
\xk{-\frac{2}{\beta_-}\frac{1-w}{1+w}}. \eea This result is
consistent with that presented in Ref.~\cite{PapajNonH}. In fact,
the self-energy $\Sigma_{22}\xk{\ep}$ can be solved by making
expansion in powers of $\ep$, as previously showed in
Ref.~\cite{PapajNonH}.

If only the $\s_z=+1$ orbital is disordered, the solutions are \bea
\Ga_+=2\La\sqrt{\frac{1+w}{1-w}} \exp\xk{-\frac{2}{\beta_+}
\frac{1+w}{1-w}}, \Ga_-=0. \eea This result is already analyzed in
the main text. Here, $\Ga_+$ represents a low energy scale below
which the perturbative RG becomes invalid. Similarly, the
self-energy can be solved by power expansion at small $\ep$ in two
limits, namely $\abs{\ep} \gg \Ga_+ $ and $\abs{\ep} \ll \Ga_+ $.
According to \Eq{EqrelaA}, we find that $\Sigma_{22}=0$ due to
$\beta_-=\beta_x=\beta_y=0$, and the left hand side can be obtained
by setting $\Sigma_{22}=\beta_x=\beta_y=0$ in \Eq{EqCoSCBA11}. Now
we obtain \begin{widetext} \bea \Sigma_{11}(\ep)&=&\frac{2\ep}{1-w}
\pm 2i\La \sqrt{\frac{1+w}{1-w}}
\exp\dk{\frac{-2\Sigma_{11}(\ep)\xk{1+w}}{\beta_+
\zk{\xk{1-w}\Sigma_{11}(\ep)-2\ep}}}. \label{EqCoSCBA11+} \eea In
the small energy regime, one can show that \bea \Sigma_{11}^R=
-\dfrac{4(1+w)\ep}{(1-w)^2\beta_+}-i\Gamma_+, \quad
\xk{|\ep|\ll\Gamma_0}.\label{EqSi11RS} \eea In the limit of
$\abs{\ep} \gg \Ga_+ $, we know from \Eq{EqCoSCBA11+} that \bea
\Sigma_{11}\xk{\ep}=\frac{2\ep}{1-w}+\frac{\beta_+\xk{1-w}^2
\zk{\Sigma_{11}\xk{\ep}-\frac{2\ep}{1-w}}\ln\zk{\Sigma_{11}\xk{\ep}
-\frac{2\ep}{1-w}}+4\ep\xk{1+w}}{\beta_+\xk{1-w}^2 \zk{\ln\xk{\La
\sqrt{\frac{1+w}{1-w}}}-\frac{i\pi}{2}}-2\xk{1-w^2}}. \eea We then
iterate the above equation once, and perform an expansion in powers
of $\beta_+$, which leads to \bea \Sigma_{11}^R\xk{\ep} =
\frac{-2\beta_+}{1+w}\zk{\ep\ln\xk{\frac{\La\sqrt{1-w^2}}{\ep}}
+\frac{i\pi\abs{\ep}}{2}}, \xk{|\ep|\gg\Gamma_0}.
\label{EqSi11RL} \eea Making use of \Eq{EqSi11RS} and \Eq{EqSi11RL},
we finally get \Eq{Eqsigma11s} of the main text.\end{widetext}

\end{document}